\documentclass[journal]{IEEEtran}
\IEEEoverridecommandlockouts
\usepackage{amsthm}

\usepackage{booktabs}
\usepackage{threeparttable}
\usepackage{amsmath,graphicx,cite,amssymb}
\usepackage{amsfonts,multirow,bm,array,setspace,stfloats}
\usepackage{graphicx,float,cite,amssymb}
\usepackage{graphics}
\DeclareGraphicsExtensions{.pdf,.jpeg,.png,.jpg}   
\usepackage{multirow,bm,bbm,array,setspace}
\usepackage{textcomp}
\usepackage{multirow}
\usepackage{diagbox}
\usepackage{color}
\usepackage{bbm}
\usepackage{tabularx}
\usepackage{array}

\usepackage{graphicx}
\usepackage{subfigure}
\usepackage{graphicx}
\usepackage{makecell}
\usepackage{enumitem}
\usepackage[normalem]{ulem}
\usepackage[caption=false,font=normalsize,labelfont=sf,textfont=sf]{subfig}
\usepackage{algorithm,algpseudocode}

\usepackage{mathrsfs}
\usepackage{ulem}

\definecolor{semanticblue}{RGB}{0,102,204}
\definecolor{errorred}{RGB}{210,0,0}
\begin{document}

\title{Single-RF-Chain Multiuser Semantic Communications via Metasurface Modulation}

\author{
 \IEEEauthorblockN{
 Mingxiao Li, \IEEEmembership{Graduate Student Member,~IEEE}, Wenhai Lai, \IEEEmembership{Member,~IEEE},\\ Hei Victor Cheng, \IEEEmembership{Member,~IEEE}, and Kaiming Shen, \IEEEmembership{Senior Member,~IEEE}
 } 
 \thanks{Mingxiao Li and Wenhai Lai contributed equally to this work.
 
 Mingxiao Li and Kaiming Shen are with the School of Science and Engineering, The Chinese University of Hong Kong, Shenzhen, China (emails: mingxiaoli@link.cuhk.edu.cn;shenkaiming@cuhk.edu.cn).

Wenhai Lai is with the Queensland Technology International School of Engineering, Dalian University of Technology, China (e-mail: wenhailai@dlut.edu.cn).

 Hei Victor Cheng is with the Department of Electrical and Computer Engineering, Aarhus University, Denmark (email: hvc@ece.au.dk).
 }
            
 }




\maketitle

\begin{abstract}
Multiuser transmission with multiple radio-frequency (RF) chains enables flexible spatial multiplexing but incurs increased hardware complexity and power consumption. A programmable metasurface (MTS) provides an alternative means of introducing spatial degrees of freedom with a single-RF-chain access point (AP). This paper proposes \emph{Semantic Prism}, an MTS-enabled framework for concurrently transmitting independent semantic messages to multiple users. Semantic Prism jointly maps multiuser semantic representations to a common AP-symbol sequence and a corresponding sequence of discrete MTS configurations, enabling each user to recover its intended message from the received signal sequence. This mapping is realized by a learning-based modulator that takes both the multiuser semantic representations and channel state information (CSI) as inputs, while accounting for the discrete phase-control constraint of the MTS. To facilitate practical deployment, we further develop a field-channel fine-tuning method that enhances robustness to CSI uncertainty. We successfully implement Semantic Prism on a 400-element MTS prototype with 2-bit phase control operating at 3.5-GHz. Field tests show that Semantic Prism substantially improves multiuser semantic reconstruction over the considered baselines and enables a single-RF-chain AP to concurrently serve up to seven users.

\end{abstract}

\section{Introduction}

A programmable metasurface (MTS) provides flexible control over wireless propagation~\cite{wu2019intelligent}. In parallel, semantic communication focuses on preserving meaning rather than exact bit recovery~\cite{xie2021deep}. Building on information-bearing MTS designs, we investigate the concurrent delivery of independent semantic messages from an access point (AP) with a single radio-frequency (RF) chain to multiple users. To this end, we propose \emph{Semantic Prism}, which jointly maps multiuser semantic representations and channel state information (CSI) to a common AP symbol sequence and a corresponding sequence of discrete MTS phase configurations. Consequently, each user's received signal sequence conveys the semantic representation intended for that user. A user-specific semantic decoder then reconstructs the corresponding message from this representation.

\begin{figure}[t]
    \centering
    \includegraphics[width=\linewidth]{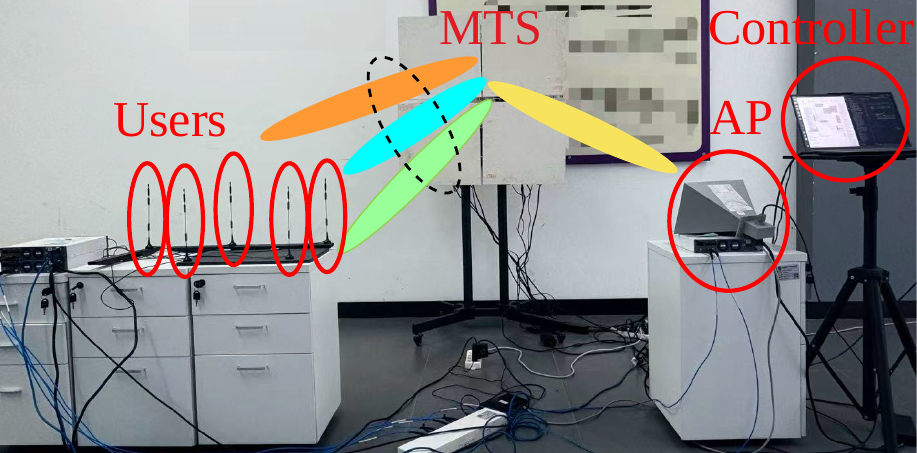}
    \caption{Overview of Semantic Prism. Unlike conventional MTS-assisted systems, Semantic Prism jointly designs AP symbols and MTS phases for modulation, enabling a single-RF-chain AP to simultaneously serve multiple users with different semantic messages.}
    \label{fig:system}
\end{figure}

An MTS-assisted single-RF-chain architecture is attractive for multiuser transmission because it can exploit the spatial degrees of freedom provided by a programmable surface without requiring multiple active RF chains, thereby reducing hardware complexity and power consumption~\cite{bereyhi2020singleRF,li2021singleRF}. However, reliable multiuser multiplexing remains challenging because existing symbol-level methods often rely on restrictive array and channel assumptions. For example, spatial Sigma--Delta ($\Sigma\Delta$) modulation maps a continuous zero-forcing (ZF) precoder to discrete constant-modulus MTS coefficients while shaping the phase quantization errors away from the angular sector occupied by the users~\cite{keung2024transmitting}. Its effectiveness depends on a one-dimensional uniform array, a steering-vector-based channel model, and sufficiently accurate CSI. In practical deployments, violations of these assumptions, together with hardware impairments, can substantially degrade symbol-level recovery. Leveraging Transformer-based models~\cite{xie2021deep,vaswani2017attention}, semantic communication offers a more robust alternative by preserving the intended meaning rather than requiring exact symbol recovery.

Most existing studies treat MTS as an auxiliary device that reshapes the propagation channel to improve multiuser transmission, coverage, 
connectivity, or physical-layer security 
\cite{huang2019RIS,wu2020ris,wu2019intelligent,cui2019secure,chen2021pushing,chen2023seamless}. In such systems, information is generated entirely by the active transmitter, and the MTS configuration is used only for channel enhancement. A more closely related line of work makes the MTS states message-dependent, allowing the surface itself to contribute directly to information transmission. Specifically, MTS states have been used to jointly encode information and shape the channel \cite{roy2021ris}, embed additional information into passive beamforming patterns \cite{yan2020ris}, map data symbols onto reflection states through reflecting modulation \cite{guo2020ris}, and directly synthesize desired constellation symbols at the receivers through symbol-level phase control \cite{cheng2022ris}. Another study has characterized the additional degrees of freedom offered by MTS phase modulation \cite{cheng2023degree}. Building on this information-bearing MTS paradigm, Semantic Prism considers message-dependent modulation for multiuser semantic transmission.

Existing semantic communication systems have been developed for the transmission of text \cite{xie2021deep}, speech \cite{weng2021semantic}, and images \cite{bourtsoulatze2019deep}; subsequent studies have considered lightweight distributed architectures for Internet of Things (IoT) devices \cite{xie2021lite}, unified multitask and multimodal transmission \cite{zhang2024unified}, and task-oriented multiuser communication \cite{xie2022task,shi2023task}. Semantic-aware resource allocation has also been investigated, such as energy-efficient semantic transmission with rate splitting \cite{yang2023energy}. Recent studies have also explored MTS-assisted semantic communication \cite{jiang2024ris,xie2024star,ma2024enhanced}. Earlier work considered MTS-assisted text semantic transmission under fading and CSI uncertainty \cite{shi2023ris}, while quality-of-experience (QoE)-aware MTS optimization was developed to jointly enhance semantic and communication performance \cite{hu2024ris}. However, these schemes still use the MTS primarily for channel enhancement. In contrast, Semantic Prism jointly maps multiuser semantic features to a common AP-symbol sequence and a corresponding sequence of MTS configurations, enabling concurrent transmission of independent semantic messages to multiple users with a single RF chain. Our main contributions are threefold:

\begin{itemize}
    \item \textbf{Single-RF-chain multiuser semantic transmission via MTS modulation.}
    We propose Semantic Prism, a semantic modulation framework that enables a single-RF-chain AP to concurrently transmit independent semantic messages to multiple users. Unlike conventional MTS-assisted systems that use the surface solely for channel enhancement, Semantic Prism makes the MTS configuration information-dependent and jointly designs a common AP-symbol sequence and a corresponding sequence of discrete MTS configurations to convey user-specific semantic representations.

    \item \textbf{CSI-aware modulation and robust field fine-tuning.}
    We propose a CSI-aware modulator that jointly maps multiuser semantic representations and CSI to AP symbols and discrete MTS configurations, with the semantic transceiver is trained end to end. We further develop a robust field fine-tuning method that incorporates perturbations of the estimated direct and reflected channels to improve robustness to CSI uncertainty in practical channels.

    \item \textbf{Prototype implementation and real-world validation.}
    We implement Semantic Prism on a 3.5-GHz prototype comprising a 400-element programmable MTS with 2-bit phase control. Field tests demonstrate substantial gains over baselines that combine separated source--channel coding with $\Sigma\Delta$ modulation. Using a single RF chain, the prototype concurrently serves up to seven users, while its aggregate effective throughput increases with the number of users, demonstrating a practical semantic multiplexing gain. Simulation results further confirm the performance advantage of Semantic Prism over the considered baselines under various system settings.
\end{itemize}

The remainder of this paper is organized as follows. Section~\ref{sec:sys_model} presents the system model and formulates the problem of concurrently transmitting independent semantic messages. Section~\ref{sec:proposed_method} develops the encoder, CSI-aware modulator, user-specific decoders, and end-to-end training procedure, and reviews the spatial $\Sigma\Delta$ precoding used in the separated source--channel coding baselines. Section~\ref{sec:channel_est} describes the channel-estimation method and robust field-channel fine-tuning approach for practical deployment. Section~\ref{sec:experiments} reports the field test and simulation results. Finally, Section~\ref{sec:conclusion} concludes the paper and outlines future research directions.

\emph{Notation:} Scalars are denoted by italic letters, while vectors and matrices are denoted by bold lowercase and bold uppercase letters, respectively. The sets of real and complex numbers are denoted by $\mathbb{R}$ and $\mathbb{C}$. For a complex scalar $u$, $\mathfrak{Re}\{u\}$ and $\mathfrak{Im}\{u\}$ denote its real part and imaginary part, respectively. For a vector $\mathbf a$, $\mathbf a^{\mathsf T}$ and $\mathbf a^{\mathsf H}$ denote its transpose and Hermitian transpose, respectively. The Euclidean norm is denoted by $\|\cdot\|_2$.

\section{System Model and Problem Formulation}
\label{sec:sys_model}

\begin{figure*}[t]
    \centering
    \includegraphics[width=0.9\linewidth]{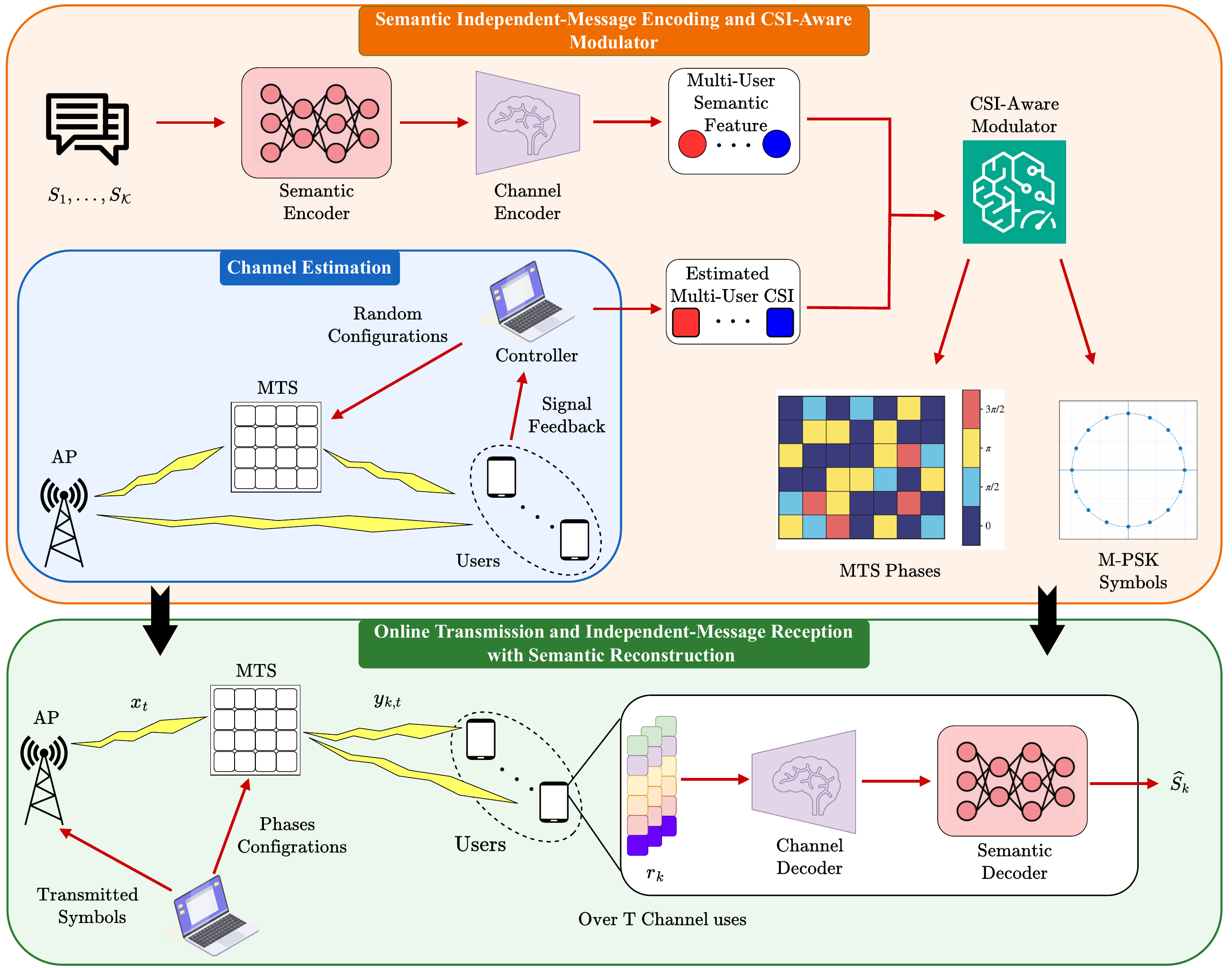}
    \caption{Workflow of Semantic Prism. The system operates in three phases:
    (i) channel estimation, in which the controller acquires the CSI;
    (ii) semantic encoding and CSI-aware modulation, in which the controller generates the AP transmit symbols and MTS configurations based on the multiuser semantic features and the estimated CSI; and
    (iii) online transmission and semantic reconstruction, in which each user reconstructs its intended message from its received symbol sequence.
    The AP transmit symbol and the MTS configuration are generated for each channel use based on the corresponding semantic features and the available CSI.}
    \label{fig:workflow}
\end{figure*}

We consider an MTS-assisted multiuser downlink semantic
communication system, as illustrated in Fig.~\ref{fig:system}, in which a single-RF-chain AP serves $K$ users with the aid of an MTS comprising $N$ programmable meta-atoms. The overall workflow of Semantic Prism is summarized in Fig.~\ref{fig:workflow}. In the system model, the direct AP--user links are assumed to be blocked, and all transmissions therefore rely on the cascaded AP--MTS--user links.

Let $\mathcal{K}\triangleq\{1,2,\ldots,K\}$ denote the user set. For user $k\in\mathcal{K}$, let $h_{n,k}\in\mathbb{C}$ denote the cascaded channel coefficient associated with the AP--MTS--user path through the $n$th meta-atom. The cascaded channel vector of user $k$ is defined as
\begin{equation}
\label{eq:reflected_channel}
\mathbf{h}_k
=
[h_{1,k},h_{2,k},\ldots,h_{N,k}]^{\mathsf T}
\in\mathbb{C}^{N}.
\end{equation}

At channel use $t$, the $n$th meta-atom applies a unit-modulus
reflection coefficient $e^{j\theta_{n,t}}$ to the incident signal.
Because each meta-atom is controlled by $B$ bits, it supports
$M=2^B$ discrete phase states. Accordingly, the phase $\theta_{n,t}$
must belong to
\begin{equation}
\left\{
0,\frac{2\pi}{M},2\frac{2\pi}{M},\ldots,
(M-1)\frac{2\pi}{M}
\right\}.
\end{equation}
Equivalently, the reflection coefficient must be selected from the
finite alphabet
\begin{equation}
\label{eq:phase_alphabet}
\Phi
=
\left\{
e^{j\frac{2\pi m}{M}}
\,\middle|\,
m=0,1,\ldots,M-1
\right\}.
\end{equation}
The MTS reflection vector at channel use $t$ is defined as
\begin{equation}
\boldsymbol{\rho}_t
=
\left[
e^{j\theta_{1,t}},
e^{j\theta_{2,t}},
\ldots,
e^{j\theta_{N,t}}
\right]^{\mathsf T}
\in\Phi^N.
\end{equation}

The AP aims to deliver an independent user-specific message
$\mathbf{S}_k$ to each user $k$. In particular, the message intended for
user $k$ consists of $J_k$ sentences and is represented as
\begin{equation}
\mathbf{S}_k
=
\{S_{k,1},S_{k,2},\ldots,S_{k,J_k}\}.
\end{equation}
The messages intended for different users are assumed to be mutually independent.

Throughout the paper, we assume block fading, such that the channel $\mathbf h_k$ remains unchanged over a transmission block of $T$ channel uses and is perfectly known to the centralized controller. Within each block, the AP transmits the complex-symbol sequence $\{x_t\}_{t=1}^{T}$, while the MTS applies the corresponding reflection-vector sequence $\{\boldsymbol{\rho}_t\}_{t=1}^{T}$. Practical CSI acquisition and imperfect CSI are considered in Section~\ref{sec:channel_est}. At channel use $t$, the signal received by user $k$ is
\begin{equation}
\label{eq:rx_signal_sys}
\begin{aligned}
y_{k,t}
&=
\left(
\sum_{n=1}^{N}
h_{n,k}e^{j\theta_{n,t}}
\right)x_t+n_{k,t},
\end{aligned}
\end{equation}
where $n_{k,t}\sim\mathcal{CN}(0,\sigma^2)$ denotes additive white
Gaussian noise.

By stacking the received symbols over the $T$ channel uses, the
received signal sequence of user $k$ is given by
\begin{equation}
\mathbf{y}_k
=
[y_{k,1},y_{k,2},\ldots,y_{k,T}]^{\mathsf T}.
\end{equation}
Based on $\mathbf{y}_k$, user $k$ reconstructs its intended message as
\begin{equation}
\widehat{\mathbf{S}}_k
=
\{
\widehat{S}_{k,1},
\widehat{S}_{k,2},
\ldots,
\widehat{S}_{k,J_k}
\}.
\end{equation}

Our objective is to jointly design the semantic transmission strategy
and the MTS reflection configurations so as to preserve the semantic
content of the source messages in their reconstructions. To quantify
the semantic fidelity of the reconstructed messages, we adopt a
sentence-level semantic similarity metric for text transmission
\cite{li2025semantic}. Specifically, for each source and reconstructed
sentence pair $(S_{k,j},\widehat{S}_{k,j})$, the adopted Sentence-BERT
model~\cite{reimers2019sentence} maps the two sentences into
$384$-dimensional real-valued embedding vectors, denoted by
$\boldsymbol{\mu}_{k,j}\in\mathbb{R}^{384}$ and
$\widehat{\boldsymbol{\mu}}_{k,j}\in\mathbb{R}^{384}$, respectively.
The Sentence-BERT-based semantic similarity~\cite{rahutomo2012semantic, crossword} is given by
\begin{equation}
\label{eq:sentence_semantic_similarity}
\lambda(S_{k,j},\widehat{S}_{k,j})
=
\frac{
\boldsymbol{\mu}_{k,j}^{\mathsf T}
\widehat{\boldsymbol{\mu}}_{k,j}
}{
\|\boldsymbol{\mu}_{k,j}\|_2
\|\widehat{\boldsymbol{\mu}}_{k,j}\|_2
}.
\end{equation}
Accordingly, the semantic quality
achieved by user $k$ over its complete message is defined as the
average similarity between its source and reconstructed sentences
\begin{equation}
\label{eq:semantic_quality}
\mathcal{Q}(\mathbf{s}_k,\widehat{\mathbf{s}}_k)
\triangleq
\frac{1}{J_k}
\sum_{j=1}^{J_k}
\lambda(S_{k,j},\widehat{S}_{k,j}).
\end{equation}

For transmission design, we consider one semantic transmission block,
in which one sentence is delivered to each user. Let $S_k$ and
$\widehat S_k$ denote the source and reconstructed sentences of user
$k$ in the current block. Over $T$ channel uses, our objective is to
jointly design the AP transmit symbols and the MTS reflection
configurations to maximize the average semantic similarity across all
users. The corresponding problem is formulated as
\begin{equation}
\label{eq:problem_formulation}
\begin{aligned}
\underset{\{x_t,\boldsymbol{\rho}_t\}_{t=1}^{T}}{\rm maximize}
\quad &
\frac{1}{K}
\sum_{k=1}^{K}
\lambda(S_k,\widehat S_k)
\\
\text{subject to}\quad
&
\,e^{j\theta_{n,t}}\in\Phi,
\quad
\forall n,t
\\
&
\frac{1}{T}
\sum_{t=1}^{T}
|x_t|^2
\leq P,
\end{aligned}
\end{equation}
where $P$ denotes the transmit-power budget and $T$ denotes the number of channel uses associated with one semantic transmission block. In \eqref{eq:problem_formulation}, the AP symbols and MTS configurations are jointly generated from the multiuser source sentences and CSI, while each user reconstructs its intended sentence from the received signal. Since the discrete MTS states and nonlinear semantic reconstruction make direct optimization difficult, we develop the end-to-end learning framework in Section~\ref{sec:proposed_method} and jointly train the encoder, CSI-aware modulator, and user-specific decoders using a token-level cross-entropy loss.

\section{Proposed Method}
\label{sec:proposed_method}

This section develops Semantic Prism, an end-to-end framework for concurrent multiuser semantic transmission with a single-RF-chain AP and an MTS. It consists of a shared semantic encoder, a CSI-aware modulator, and user-specific decoders, which are jointly trained end to end. We further review the spatial $\Sigma\Delta$ precoding used in the separated source--channel coding baselines and highlight its distinction from Semantic Prism.

\subsection{Semantic Encoding of Independent Messages}
\label{subsec:semantic_encoding}

Each semantic transmission block carries one sentence for each user.
Consider the $j$th sentence $S_{k,j}$ in the message $\mathbf s_k$
intended for user $k$. Since all sentences are processed in the same
manner, we omit the sentence index $j$ and write
\begin{equation}
S_k \triangleq S_{k,j}.
\end{equation}
A message containing multiple sentences is transmitted over multiple
semantic transmission blocks. After tokenization, $S_k$ is represented by
\begin{equation}
\label{eq:token_sequence}
\mathbf w_k
=
[w_{k,1},w_{k,2},\ldots,w_{k,L}],
\qquad
w_{k,\ell}\in\mathcal V,
\end{equation}
where $L$ is the maximum sequence length and
$\mathcal V=\{v_0,v_1,\ldots,v_{|\mathcal V|-1}\}$ denotes the
vocabulary. Sentences shorter than $L$ are padded, and the padding
positions are excluded from the training loss.

Each token $w_{k,\ell}$ is represented by a one-hot vector
$\mathbf e_{k,\ell}\in\{0,1\}^{|\mathcal V|}$. A shared trainable
embedding matrix $\mathbf E\in\mathbb R^{d\times|\mathcal V|}$ then
maps it to a $d$-dimensional representation as
\begin{equation}
    \mathbf f_{k,\ell}
    =
    \mathbf E\mathbf e_{k,\ell}
    \in\mathbb R^d.
\end{equation}
Stacking the token embeddings gives
\begin{equation}
\label{eq:transformer_input}
\mathbf F_k
=
\left[
\mathbf f_{k,1},
\mathbf f_{k,2},
\ldots,
\mathbf f_{k,L}
\right]^{\mathsf T}
\in\mathbb R^{L\times d}.
\end{equation}

The shared Transformer-based~\cite{vaswani2017attention} semantic encoder maps $\mathbf F_k$ to the contextual semantic representation
\begin{equation}
\label{eq:semantic_feature}
\mathbf Z_k
=
\operatorname{TE}_{\boldsymbol\varphi}(\mathbf F_k)
\in\mathbb R^{L\times d},
\end{equation}
where $\boldsymbol\varphi$ denotes the encoder parameters. Let
$\mathbf z_{k,\ell}\in\mathbb R^d$ denote the $\ell$th row of
$\mathbf Z_k$.

To reduce the dimensionality of the semantic features before physical
transmission, a shared channel encoder maps each
$\mathbf z_{k,\ell}$ to
\begin{equation}
\label{eq:compressed_semantic_feature}
\mathbf u_{k,\ell}
=
\operatorname{AE}_{\boldsymbol\varphi}
(\mathbf z_{k,\ell})
\in\mathbb R^{d_u}.
\end{equation}
We set $d_u=2N_c$, where $N_c$ denotes the number of complex channel uses associated with each token. Thus, $\mathbf u_{k,\ell}$ contains $2N_c$ real-valued features corresponding to $N_c$ complex channel uses. Stacking the compressed features over all token positions yields
\begin{equation}
\label{eq:compressed_feature_sequence}
\mathbf U_k
=
\left[
\mathbf u_{k,1},
\mathbf u_{k,2},
\ldots,
\mathbf u_{k,L}
\right]^{\mathsf T}
\in\mathbb R^{L\times 2N_c}.
\end{equation}
The compressed semantic representations of all users are subsequently
combined with the CSI and processed by the CSI-aware modulator
described in the next subsection.

\subsection{CSI-Aware Modulator}
\label{subsec:csi_aware_modulator}

The CSI-aware modulator jointly maps the compressed semantic
representations of all users and the available CSI to a common AP
transmit-symbol sequence and a sequence of message-dependent MTS
reflection configurations. Recall that
$\mathbf u_{k,\ell}\in\mathbb R^{2N_c}$
denotes the compressed semantic feature associated with the
$\ell$th token of user $k$. The $(2i-1)$th and $2i$th entries of
$\mathbf u_{k,\ell}$ correspond to the in-phase and quadrature
features associated with the $i$th channel use of that token,
respectively. The channel-level semantic features of all users are
therefore concatenated as
\begin{equation}
\label{eq:multiuser_semantic_feature}
\widetilde{\mathbf u}_{\ell,i}
=
\left[
[\mathbf u_{1,\ell}]_{2i-1},
[\mathbf u_{1,\ell}]_{2i},
\ldots,
[\mathbf u_{K,\ell}]_{2i-1},
[\mathbf u_{K,\ell}]_{2i}
\right]^{\mathsf T},
\end{equation}
where
$\widetilde{\mathbf u}_{\ell,i}\in\mathbb R^{2K}$, $\ell=1,2,\ldots,L$ and $i=1,2,\ldots,N_c$.

The reflected-link CSI is converted into a real-valued feature vector
by stacking the real and imaginary parts of the reflected channels of
all users
\begin{equation}
\label{eq:reflected_csi_feature}
\mathbf c_r
=
\left[
\Re\{\mathbf h_1\}^{\mathsf T},
\Im\{\mathbf h_1\}^{\mathsf T},
\ldots,
\Re\{\mathbf h_K\}^{\mathsf T},
\Im\{\mathbf h_K\}^{\mathsf T}
\right]^{\mathsf T},
\end{equation}
where $\mathbf c_r\in\mathbb R^{2KN}$. Under the reflected-link-only model, the reserved direct-link feature is set to $\mathbf c_d = \mathbf 0_{2K}$. The semantic and CSI features are then concatenated as
\begin{equation}
\label{eq:modulator_input}
\mathbf a_{\ell,i}
=
\left[
\widetilde{\mathbf u}_{\ell,i}^{\mathsf T},
\mathbf c_d^{\mathsf T},
\mathbf c_r^{\mathsf T}
\right]^{\mathsf T}
\in
\mathbb R^{4K+2KN}.
\end{equation}

Each token is represented by $N_c$ physical channel uses. Hence, one
sentence-transmission block comprises
\begin{equation}
\label{eq:number_channel_uses}
T=LN_c
\end{equation}
channel uses. The physical channel-use index associated with token
position $\ell$ and within-token index $i$ is
\begin{equation}
\label{eq:channel_use_index}
t=(\ell-1)N_c+i.
\end{equation}

Under the block-fading assumption, the CSI remains unchanged over one semantic transmission block and is concatenated with the channel-level semantic features at every channel use. Using
\eqref{eq:channel_use_index}, let
$\mathbf a_t\triangleq\mathbf a_{\ell,i}$ and collect the modulator
inputs over the block as
\begin{equation}
\label{eq:modulator_input_sequence}
\mathbf V
=
\left[
\mathbf a_1,
\mathbf a_2,
\ldots,
\mathbf a_T
\right]^{\mathsf T}
\in
\mathbb R^{T\times(4K+2KN)}.
\end{equation}

The sequence $\mathbf V$ is processed by a bidirectional long
short-term memory (BiLSTM) network~\cite{graves2005framewise} to
exploit the sequential correlations among the channel-level semantic
features over the transmission block. Specifically,
\begin{equation}
\label{eq:bilstm_modulator}
\left[
\mathbf m_1,
\mathbf m_2,
\ldots,
\mathbf m_T
\right]^{\mathsf T}
=
\operatorname{LN}
\left(
\operatorname{BiLSTM}_{\boldsymbol\psi}(\mathbf V)
\right),
\end{equation}
where
$\mathbf m_t\in\mathbb R^{d_m}$
denotes the resulting feature at channel use $t$,
$\operatorname{LN}(\cdot)$ denotes layer normalization, and
$\boldsymbol\psi$ denotes the trainable parameters of the CSI-aware modulator. Each $\mathbf m_t$ is subsequently used to generate the AP symbol and MTS configuration for the corresponding channel use.

Two output heads operate on $\mathbf m_t$. The first generates the
common AP transmit symbol, while the second determines the discrete
reflection state of every MTS meta-atom. For the continuous-valued AP signaling, the AP-symbol head produces
\begin{equation}
\label{eq:continuous_ap_symbol_head}
\mathbf g_t^{x}
=
\operatorname{MLP}_{x,\boldsymbol\psi}
(\mathbf m_t)
\in\mathbb R^2,
\end{equation}
where
$\operatorname{MLP}_{x,\boldsymbol\psi}(\cdot)$
consists of a fully connected hidden layer with a ReLU activation,
followed by a linear output layer. The two outputs represent the
in-phase and quadrature components of an unnormalized complex symbol:
\begin{align}
\Re\{\widetilde x_t\}
&=
[\mathbf g_t^{x}]_1,
\label{eq:continuous_ap_real}
\\
\Im\{\widetilde x_t\}
&=
[\mathbf g_t^{x}]_2.
\label{eq:continuous_ap_imag}
\end{align}
Equivalently,
\begin{equation}
\label{eq:unnormalized_ap_symbol}
\widetilde x_t
=
[\mathbf g_t^{x}]_1
+
j[\mathbf g_t^{x}]_2.
\end{equation}
The complete symbol sequence is normalized over the transmission block
to satisfy the average transmit-power constraint
\begin{equation}
\label{eq:ap_power_normalization}
x_t
=
\frac{
\sqrt{P}\,\widetilde x_t
}{
\sqrt{
\frac{1}{T}
\sum_{\tau=1}^{T}
|\widetilde x_\tau|^2
}
},
\qquad
t=1,2,\ldots,T.
\end{equation}

The MTS-control head generates the logits associated with all
candidate reflection states
\begin{equation}
\label{eq:mts_control_logits}
\mathbf g_t^{\theta}
=
\operatorname{MLP}_{\theta,\boldsymbol\psi}
(\mathbf m_t)
\in\mathbb R^{NM},
\end{equation}
where
$\operatorname{MLP}_{\theta,\boldsymbol\psi}(\cdot)$
consists of a fully connected hidden layer with a ReLU activation and
a linear output layer. The output is reshaped into an
$N\times M$ logit matrix. Let
\begin{equation}
\label{eq:mts_element_logits}
\mathbf z_{t,n}^{\theta}
\in\mathbb R^M
\end{equation}
denote the logits over the $M$ candidate phase states of the $n$th
meta-atom at channel use $t$.

During training, the straight-through Gumbel--Softmax
estimator~\cite{jang2017categorical} is used to enable
backpropagation through the discrete state-selection operation. Let
\begin{equation}
\label{eq:mts_one_hot_vector}
\boldsymbol{\alpha}_{t,n}
=
\left[
\alpha_{t,n,0},
\alpha_{t,n,1},
\ldots,
\alpha_{t,n,M-1}
\right]^{\mathsf T}
\in\{0,1\}^{M}
\end{equation}
denote the selected one-hot state, subject to
\begin{equation}
\label{eq:one_hot_mts_constraint}
\sum_{q=0}^{M-1}
\alpha_{t,n,q}
=
1.
\end{equation}
A hard one-hot state is used in the forward pass, whereas gradients
are propagated through the corresponding soft Gumbel--Softmax
relaxation in the backward pass.

The reflection coefficient of the $n$th meta-atom at channel use $t$
is then constructed as
\begin{equation}
\label{eq:mts_reflection_coefficient}
e^{j\theta_{n,t}}
=
\sum_{q=0}^{M-1}
\alpha_{t,n,q}
e^{j2\pi q/M}.
\end{equation}
Accordingly, the MTS reflection vector is
\begin{equation}
\label{eq:mts_reflection_vector_token}
\boldsymbol{\rho}_t
=
\left[
e^{j\theta_{1,t}},
e^{j\theta_{2,t}},
\ldots,
e^{j\theta_{N,t}}
\right]^{\mathsf T}
\in\Phi^N.
\end{equation}

During inference, no Gumbel perturbation is introduced. Instead, the
phase state of each meta-atom is selected deterministically as
\begin{equation}
\label{eq:mts_phase_argmax}
q_{t,n}^{\star}
=
\arg\max_{q\in\{0,1,\ldots,M-1\}}
[\mathbf z_{t,n}^{\theta}]_q,
\end{equation}
which yields
\begin{equation}
\label{eq:mts_phase_inference}
e^{j\theta_{n,t}}
=
e^{j2\pi q_{t,n}^{\star}/M}.
\end{equation}

Therefore, at every physical channel use, the CSI-aware modulator
generates one common AP transmit symbol $x_t$ and one
message-dependent MTS reflection configuration
$\boldsymbol{\rho}_t$. These two outputs are jointly optimized
through the end-to-end semantic communication pipeline.

\textit{Remark 1 (Discrete AP Signaling for Field Tests):}
The formulation above permits continuous-valued AP symbols. In the prototype, the AP symbols are restricted to an $M_x$-PSK alphabet to provide constant-envelope signaling and to
match the implemented RF transmission scheme. The AP constellation
order $M_x$ is independent of the number $M$ of discrete MTS phase
states. The BiLSTM backbone in \eqref{eq:bilstm_modulator} and the
MTS-control head remain unchanged; only the output layer of the
AP-symbol head is replaced.

Specifically, the AP-symbol head produces
\begin{equation}
\label{eq:discrete_ap_symbol_logits}
\mathbf g_t^{x}
=
\operatorname{MLP}_{x,\boldsymbol\psi}(\mathbf m_t)
\in\mathbb R^{M_x},
\end{equation}
where $\mathbf g_t^{x}$ contains the logits over the $M_x$ candidate
PSK symbols. During training, the straight-through Gumbel--Softmax
estimator produces the one-hot selection vector
\begin{equation}
\label{eq:ap_one_hot_vector}
\boldsymbol{\beta}_t
=
\left[
\beta_{t,0},
\beta_{t,1},
\ldots,
\beta_{t,M_x-1}
\right]^{\mathsf T}
\in\{0,1\}^{M_x},
\end{equation}
subject to
\begin{equation}
\label{eq:ap_one_hot_constraint}
\sum_{p=0}^{M_x-1}
\beta_{t,p}
=
1.
\end{equation}
The selected unit-modulus symbol is
\begin{equation}
\label{eq:discrete_ap_symbol}
\widetilde x_t
=
\sum_{p=0}^{M_x-1}
\beta_{t,p}
e^{j2\pi p/M_x}.
\end{equation}
Since $|\widetilde x_t|=1$, the transmitted symbol is
\begin{equation}
\label{eq:discrete_ap_power_scaling}
x_t
=
\sqrt{P}\,\widetilde x_t.
\end{equation}
During inference and field transmission, the AP constellation point is
selected by applying a hard $\arg\max$ operation to $\mathbf g_t^{x}$,
without introducing Gumbel perturbations.

\subsection{Independent-Message Reception and Semantic Reconstruction}
\label{subsec:semantic_reconstruction}

For user $k$, the received signal at channel use $t$ follows
\eqref{eq:rx_signal_sys}. Since each token is represented by $N_c$
channel uses, the received symbols associated with token position
$\ell$ are grouped as
\begin{equation}
\label{eq:received_symbol_block}
\mathbf y_{k,\ell}
=
\left[
y_{k,(\ell-1)N_c+1},
\ldots,
y_{k,\ell N_c}
\right]^{\mathsf T}
\in\mathbb C^{N_c}.
\end{equation}
Their in-phase and quadrature components are interleaved to form the
real-valued channel feature
$\mathbf r_{k,\ell}\in\mathbb R^{2N_c}$, where
\begin{align}
[\mathbf r_{k,\ell}]_{2i-1}
&=
\Re\left\{
y_{k,(\ell-1)N_c+i}
\right\},
\\
[\mathbf r_{k,\ell}]_{2i}
&=
\Im\left\{
y_{k,(\ell-1)N_c+i}
\right\},
\end{align}
for $i=1,\ldots,N_c$.

A user-specific feature decoder maps the received channel feature back
to the semantic-feature space
\begin{equation}
\label{eq:semantic_feature_reconstruction}
\widehat{\mathbf z}_{k,\ell}
=
\operatorname{AD}_{k,\boldsymbol\omega_k}
(\mathbf r_{k,\ell})
\in\mathbb R^d,
\end{equation}
where
$\operatorname{AD}_{k,\boldsymbol\omega_k}(\cdot)$
denotes the feature decoder of user $k$. Stacking the reconstructed
features over all token positions gives
\begin{equation}
\label{eq:reconstructed_semantic_sequence}
\widehat{\mathbf Z}_k
=
\left[
\widehat{\mathbf z}_{k,1},
\widehat{\mathbf z}_{k,2},
\ldots,
\widehat{\mathbf z}_{k,L}
\right]^{\mathsf T}
\in\mathbb R^{L\times d}.
\end{equation}

The reconstructed feature sequence is then processed by a
user-specific semantic decoder based on the transformer decoder
architecture~\cite{vaswani2017attention}:
\begin{equation}
\label{eq:transformer_decoder_output}
\boldsymbol{\Xi}_k
=
\operatorname{TD}_{k,\boldsymbol\omega_k}
(\widehat{\mathbf Z}_k)
\in\mathbb R^{L\times d},
\end{equation}
where
$\operatorname{TD}_{k,\boldsymbol\omega_k}(\cdot)$ denotes the
user-specific Qwen2-based semantic decoder, and
$\boldsymbol{\xi}_{k,\ell}\in\mathbb R^d$ denotes the $\ell$th row of
$\boldsymbol{\Xi}_k$.

The input token embedding and the output vocabulary projection share
the same trainable embedding matrix $\mathbf E$ introduced in
Section~\ref{subsec:semantic_encoding}. Accordingly, the token
posterior at position $\ell$ is computed as
\begin{equation}
\label{eq:token_posterior}
\boldsymbol{\pi}_{k,\ell}
=
\operatorname{softmax}
\left(
\mathbf E^{\mathsf T}
\boldsymbol{\xi}_{k,\ell}
\right)
\in\mathbb R^{|\mathcal V|}.
\end{equation}

During inference, the reconstructed token index is obtained by
\begin{equation}
\label{eq:reconstructed_token_index}
\widehat q_{k,\ell}
=
\arg\max_{q\in\{0,1,\ldots,|\mathcal V|-1\}}
[\boldsymbol{\pi}_{k,\ell}]_q,
\end{equation}
and the corresponding token is
\begin{equation}
\label{eq:reconstructed_token}
\widehat w_{k,\ell}
=
v_{\widehat q_{k,\ell}}.
\end{equation}
The reconstructed token sequence is therefore
\begin{equation}
\label{eq:reconstructed_token_sequence}
\widehat{\mathbf w}_k
=
\left[
\widehat w_{k,1},
\widehat w_{k,2},
\ldots,
\widehat w_{k,L}
\right].
\end{equation}
After removing padding and auxiliary tokens, the sequence is
detokenized to obtain the reconstructed sentence $\widehat S_k$.

\subsection{End-to-End Training}
\label{subsec:end_to_end_training}

During training, channel and noise realizations are generated using $\mathcal C_{\mathrm{sim}}$ based on \eqref{eq:rx_signal_sys}, and the entire semantic transceiver is jointly optimized end to end. Let
\begin{equation}
\boldsymbol{\Omega}
=
\left\{
\boldsymbol\varphi,
\boldsymbol\psi,
\boldsymbol\omega_1,
\ldots,
\boldsymbol\omega_K
\right\}
\end{equation}
denote the complete set of trainable parameters, where
$\boldsymbol\varphi$ denotes the parameters of the shared semantic and channel encoders, $\boldsymbol\psi$ contains the CSI-aware modulator parameters, and $\boldsymbol\omega_k$ contains the feature and semantic decoder parameters of user $k$.

For model training, we minimize the token-level
cross-entropy loss. Let $N_{\mathrm b}$ denote the mini-batch size,
and let
\begin{equation}
\chi_{k,\ell}^{(b)}
\in\{0,1\}
\end{equation}
denote the validity indicator for the $\ell$th token of user $k$ in
the $b$th training sample. In particular,
$\chi_{k,\ell}^{(b)}=0$ for a padding position and
$\chi_{k,\ell}^{(b)}=1$ otherwise. Using the posterior distribution in
\eqref{eq:token_posterior}, the training loss is
\begin{equation}
\label{eq:training_loss}
\mathcal L
=
-
\frac{
\displaystyle
\sum_{b=1}^{N_{\mathrm b}}
\sum_{k=1}^{K}
\sum_{\ell=1}^{L}
\chi_{k,\ell}^{(b)}
\log
\left[
\boldsymbol{\pi}_{k,\ell}^{(b)}
\right]_{\iota(w_{k,\ell}^{(b)})}
}{
\displaystyle
\sum_{b=1}^{N_{\mathrm b}}
\sum_{k=1}^{K}
\sum_{\ell=1}^{L}
\chi_{k,\ell}^{(b)}
},
\end{equation}
where $w_{k,\ell}^{(b)}$ denotes the target token and
$\iota(w_{k,\ell}^{(b)})$ denotes its vocabulary index. All trainable parameters in $\boldsymbol{\Omega}$ are optimized by minimizing \eqref{eq:training_loss} via backpropagation. The overall training procedure is summarized in Algorithm~\ref{alg:semantic_prism_training}.

\begin{algorithm}[t]
\caption{End-to-End Training of Semantic Prism}
\label{alg:semantic_prism_training}
\begin{algorithmic}[1]

\State \textbf{Input:} training corpus $\mathcal D$; user set $\mathcal K$; MTS size $N$; phase alphabet $\Phi$ of size $M$; channel uses per token $N_c$; power budget $P$; simulated channel model $\mathcal C_{\mathrm{sim}}$; mini-batch size $N_{\mathrm b}$; number of training epochs $N_{\mathrm{ep}}$.

\State Initialize $\boldsymbol{\Omega}=\{\boldsymbol\varphi,\boldsymbol\psi,\boldsymbol\omega_1,\ldots,\boldsymbol\omega_K\}$.

\For{$e=1,2,\ldots,N_{\mathrm{ep}}$}

\State Sample a mini-batch of multiuser sentences from $\mathcal D$.

\State Sample the channel and noise from $\mathcal C_{\mathrm{sim}}$.

\State Encode the token sequences to obtain $\{\mathbf Z_k^{(b)},\mathbf U_k^{(b)}\}$.

\State Generate $\{x_t^{(b)},\boldsymbol{\rho}_t^{(b)}\}_{t=1}^{T}$ via the CSI-aware modulator.

\State Obtain $\{\mathbf y_k^{(b)}\}$ through $\mathcal C_{\mathrm{sim}}$ according to \eqref{eq:rx_signal_sys}.

\State Decode $\{\mathbf y_k^{(b)}\}$ to obtain $\{\boldsymbol{\pi}_{k,\ell}^{(b)}\}$.

\State Compute the token-level loss $\mathcal L$ according to \eqref{eq:training_loss}.

\State Update $\boldsymbol{\Omega}$ by backpropagation.

\EndFor

\State \textbf{Output:} trained parameters $\boldsymbol{\Omega}$.

\end{algorithmic}
\end{algorithm}

\subsection{Spatial $\Sigma\Delta$ Precoding}
\label{sec:sigma_delta}

For completeness, we briefly review the spatial $\Sigma\Delta$ precoding \cite{keung2024transmitting} adopted in our separated source--channel coding baselines. Let $\mathbf{x}_t\in\mathbb{C}^K$ denote the symbols intended for the $K$ users at channel use $t$, and $\mathbf{H}\in\mathbb{C}^{K\times N}$ the multiuser cascaded channel matrix. With the ZF precoder $\mathbf{W}\in\mathbb{C}^{N\times K}$ satisfying $\mathbf{H}\mathbf{W}=\mathbf{I}_K$, the continuous MTS coefficient vector $\tilde{\boldsymbol{\rho}}_t=\mathbf{W}\mathbf{x}_t\in\mathbb{C}^N$ generally does not belong to the discrete phase alphabet $\Phi^N$. Spatial $\Sigma\Delta$ modulation therefore quantizes it sequentially, element by element, as
\begin{equation}
\begin{aligned}
\rho_{n,t} &= Q_\Phi\big(\tilde{\rho}_{n,t}-e^{j\phi}q_{n-1,t}\big), \\
q_{n,t} &= \rho_{n,t}-\tilde{\rho}_{n,t}+e^{j\phi}q_{n-1,t},\quad q_{0,t}=0,
\end{aligned}
\end{equation}
where $Q_\Phi(\cdot)$ denotes the nearest-phase quantizer, and the quantization error of each element is fed back when quantizing the next. Under the ULA steering-vector channel assumed in \cite{keung2024transmitting}, $h_{n+1,k}=h_{n,k}e^{-j\omega_k}$, where $\omega_k$ denotes the spatial frequency of user $k$, the residual quantization distortion at user $k$ collapses to $(1-e^{j(\phi-\omega_k)})$ times a weighted sum of the per-element error terms plus a boundary term. Hence, choosing $\phi=\omega_k$ for a single user, or $\phi$ at the center of the user angular sector for multiple users, shapes the residual quantization noise away from the intended users.

Compared with Semantic Prism, spatial $\Sigma\Delta$ precoding (i) uses a fixed AP carrier and designs only the MTS coefficients for prescribed user symbols, whereas Semantic Prism jointly designs the AP transmit symbol $x_t$ and the discrete MTS configuration $\boldsymbol{\rho}_t$ from the multiuser semantic representations and CSI, and (ii) relies on the regular phase progression of a ULA steering-vector channel for noise shaping, whereas Semantic Prism directly uses the element-wise cascaded CSI.

\section{Practical Deployment of Semantic Prism}
\label{sec:channel_est}

The preceding design assumes blocked direct AP--user links and perfect CSI. In practical environments, however, the direct links may be non-negligible and the estimated reflected CSI may contain errors. We therefore introduce a fine-tuning scheme to improve robustness.

\subsection{Channel Estimation}
\label{subsec:channel_estimation}

During channel estimation, the AP transmits a fixed known signal $x_t=\sqrt{P}$, while the MTS applies $T_{\rm ce}$ random phase-shift configurations. The direct and reflected channels are assumed to remain unchanged over the $T_{\rm ce}$ channel-estimation samples. Let
\begin{equation}
\boldsymbol{\rho}_t
=
\left[
e^{j\theta_{1,t}},
e^{j\theta_{2,t}},
\ldots,
e^{j\theta_{N,t}}
\right]^{\mathsf T},
\end{equation}
where each entry is independently and uniformly drawn from $\Phi$. For user $k$, the received pilot measurements are collected as
\begin{equation}
\mathbf y_k^{\rm ce}
=
\left[
y_{k,1}^{\rm ce},
y_{k,2}^{\rm ce},
\ldots,
y_{k,T_{\rm ce}}^{\rm ce}
\right]^{\mathsf T}.
\end{equation}
Define
\begin{equation}
\mathbf A
=
\begin{bmatrix}
1
&
\cdots
&
1
\\
e^{j\theta_{1,1}}
&
\cdots
&
e^{j\theta_{1,T_{\rm ce}}}
\\
\vdots
&
\ddots
&
\vdots
\\
e^{j\theta_{N,1}}
&
\cdots
&
e^{j\theta_{N,T_{\rm ce}}}
\end{bmatrix}
\in\mathbb C^{(N+1)\times T_{\rm ce}}.
\end{equation}
The first row accounts for the direct AP--user link, while the remaining $N$ rows correspond to the reflected links through the meta-atoms. We choose \(T_{\rm ce}\) to be larger than \(N\) and use independently generated random MTS configurations such that \(\mathbf A\) has full row rank. The received measurements satisfy
\begin{equation}
\mathbf y_k^{\rm ce}
=
\sqrt{P}\,
\mathbf A^{\mathsf T}
\begin{bmatrix}
d_k\\
\mathbf h_k
\end{bmatrix}
+
\mathbf n_k^{\rm ce},
\end{equation}
where
$\mathbf n_k^{\rm ce}\in\mathbb C^{T_{\rm ce}}$
denotes the channel-estimation noise vector. The direct and reflected
channels are jointly estimated as
\begin{equation}
\label{eq:channel_est}
\begin{bmatrix}
\widehat d_k\\
\widehat{\mathbf h}_k
\end{bmatrix}
=
\frac{1}{\sqrt{P}}
\left(
(\mathbf y_k^{\rm ce})^{\mathsf T}
\mathbf A^+
\right)^{\mathsf T}.
\end{equation}
where $\mathbf A^+$ denotes the pseudo-inverse of $\mathbf A$.
Algorithm~\ref{alg:channel_est} summarizes the procedure.

\begin{algorithm}[t]
\caption{Channel Estimation Algorithm}
\label{alg:channel_est}
\begin{algorithmic}[1]

\State \textbf{Input:} phase set $\Phi$, number of meta-atoms $N$, and number of estimation samples $T_{\rm ce}$.

\For{$t=1,2,\ldots,T_{\rm ce}$}

\State Generate each $e^{j\theta_{n,t}}$ uniformly from $\Phi$.

\State Each user feeds back its pilot measurement under $\boldsymbol{\rho}_t$.

\EndFor

\State Recover $\{\widehat d_k,\widehat{\mathbf h}_k\}_{k\in\mathcal K}$ according to \eqref{eq:channel_est}.

\State \textbf{Output:} channel estimates $\{\widehat d_k,\widehat{\mathbf h}_k\}_{k\in\mathcal K}$.

\end{algorithmic}
\end{algorithm}

\subsection{Robust Field-Channel Fine-Tuning}
\label{subsec:robust_ft}

For field deployment, we fine-tune the pretrained model using the
estimated direct and reflected channels with channel perturbations.
For user $k$, the received signal over the physical channel is
\begin{equation}
    y_{k,t}
    =
    \left(
        d_k+\mathbf h_k^{\mathsf T}\boldsymbol{\rho}_t
    \right)x_t+n_{k,t}.
    \label{eq:field_channel}
\end{equation}

The controller uses the channel estimates
$\{\widehat d_k,\widehat{\mathbf h}_k\}$ obtained in
Section~\ref{subsec:channel_estimation}. In the modulator input, the zero-valued direct-link feature $\mathbf c_d$ in \eqref{eq:modulator_input} is replaced by the real-valued representation of $\{\widehat d_k\}_{k=1}^{K}$. Similarly, the reflected-link feature $\mathbf c_r$ is constructed from
$\{\widehat{\mathbf h}_k\}_{k=1}^{K}$ according to
\eqref{eq:reflected_csi_feature}.

To improve robustness to channel-estimation errors, we augment both
the direct and reflected channel estimates during fine-tuning. Define
their powers as
\begin{equation}
    \widehat p_{d,k}
    =
    |\widehat d_k|^2,
    \qquad
    \widehat p_{r,k}
    =
    \frac{1}{N}
    \left\|
        \widehat{\mathbf h}_k
    \right\|_2^2.
    \label{eq:channel_power}
\end{equation}
The augmented channels are generated as
\begin{align}
    \widetilde d_k
    &=
    \widehat d_k+\delta_k,
    \label{eq:direct_perturb}
    \\
    \widetilde{\mathbf h}_k
    &=
    \widehat{\mathbf h}_k+\boldsymbol{\Delta}_k,
    \label{eq:reflected_perturb}
\end{align}
where
\begin{align}
    \delta_k \mid \widehat d_k
    &\sim
    \mathcal{CN}
    \left(
        0,
        \eta_{\rm csi}^2\widehat p_{d,k}
    \right),
    \\
    \boldsymbol{\Delta}_k
    \mid \widehat{\mathbf h}_k
    &\sim
    \mathcal{CN}
    \left(
        \mathbf 0,
        \eta_{\rm csi}^2
        \widehat p_{r,k}\mathbf I_N
    \right).
    \label{eq:channel_perturbation}
\end{align}
The parameter $\eta_{\rm csi}\geq0$ controls the perturbation level,
with $\eta_{\rm csi}=0$ corresponding to fine-tuning without channel
augmentation.

The modulator always uses the measured CSI
$\{\widehat d_k,\widehat{\mathbf h}_k\}$, while the perturbed channels
$\{\widetilde d_k,\widetilde{\mathbf h}_k\}$ are used only for forward propagation
\begin{equation}
    \widetilde y_{k,t}
    =
    \left(
        \widetilde d_k+
        \widetilde{\mathbf h}_k^{\mathsf T}\boldsymbol{\rho}_t
    \right)x_t+n_{k,t}.
    \label{eq:perturbed_field_channel}
\end{equation}
Thus, the model is fine-tuned under randomized channel mismatch while
the modulator operates on the measured CSI. During field transmission,
the measured CSI is used to generate the AP symbols and MTS
configurations, which are transmitted over the physical channel in
\eqref{eq:field_channel}.

\section{Experiments}
\label{sec:experiments}


\subsection{Experimental Setup}
\label{subsec:experimental_setup}

\textbf{Dataset Details.} We adopt the English Europarl corpus~\cite{koehn-2005-europarl} dataset to evaluate the proposed Semantic Prism. Specifically, the dataset comprises
approximately $80{,}000$ English sentences and $1.5$ million words. We randomly partition the whole dataset into training, validation, and test sets with an $8{:}1{:}1$ ratio. Each multiuser sample is constructed by independently drawing one sentence per user from the corresponding split. Each multi-user sample is constructed by independently drawing one sentence per user from the corresponding split. 

\textbf{MTS prototype and communication system.}
The MTS prototype used in our field tests integrates four
custom-designed $10\times10$ modules into a $20\times20$ array with
$N=400$ meta-atoms. Each meta-atom employs two PIN diodes and supports
four 2-bit phase states, $\{0,\pi/2,\pi,3\pi/2\}$. The MTS operates at
3.5-GHz with a $100$~MHz bandwidth. The single-RF-chain AP and user
nodes are implemented using NI-USRP X410 devices, with a directional
antenna at the AP and an omnidirectional antenna at each user. The
signal bandwidth is $125$~kHz, and the receivers sample the complex
baseband signals at $700$~kS/s. The transmit power is set to
$P=10$~dBm.

For each measurement state, the AP symbol and MTS configuration are
updated and then jointly held for $100$~ms. The receivers continuously acquire complex baseband samples, and the samples collected over the last $70$~ms of each holding interval are averaged to form the received observation. The received observations are aligned with the corresponding AP symbols and MTS configurations using common timestamps recorded on the same laptop. Following Section~\ref{sec:channel_est}, the direct- and reflected-link
CSI are jointly estimated from pilot measurements collected under
$T_{\rm ce}$ random MTS configurations. Unless otherwise stated, we set
$T_{\rm ce}=5000$ and use $\eta_{\rm csi}=1$ for field-channel
fine-tuning.

For reference, useful symbol durations in representative 5G New Radio (NR) numerologies are approximately $8.33$--$66.7~\mu$s
\cite{3gpp38211}. Existing MTS prototypes have demonstrated
reconfiguration speeds within this time scale. In particular,
\cite{xu2025chaotic} reports a minimum switching interval of
$2.5~\mu$s for a PIN-diode information metasurface, while
\cite{wang2025ris} demonstrates MTS pattern switching at
$12.5$~MHz, corresponding to an $80$-ns switching interval.
These results indicate that symbol-level MTS reconfiguration is
feasible with high-speed control hardware. The $100$-ms holding
interval used in our prototype is chosen for measurement purposes.


\begin{table}[t]
\centering
\caption{Network Configuration of Semantic Prism in Field Test }
\label{tab:network_configuration}
\small
\renewcommand{\arraystretch}{1.08}
\setlength{\tabcolsep}{3pt}

\begin{tabular}{c l c c}
\toprule
\textbf{Module}
& \textbf{Layer}
& \textbf{Units}
& \textbf{Activation} \\
\midrule

\multirow{3}{*}{\shortstack{Semantic\\Encoder}}
& $1\times$ Transformer
& 128
& SiLU \\
& Dense + Dropout
& 256
& SiLU \\
& Dense
& 32
& Linear \\
\midrule

\multirow{5}{*}{\shortstack{CSI-Aware\\Modulator}}
& $1\times$ BiLSTM
& 512
& -- \\
& AP-symbol Dense
& 128
& ReLU \\
& AP-symbol Output
& $M_x$
& Gumbel--Softmax \\
& MTS-control Dense
& 128
& ReLU \\
& MTS-control Output
& $NM$
& Gumbel--Softmax \\
\midrule

\multirow{4}{*}{\shortstack{Semantic\\Decoder}}
& Dense + Dropout
& 256
& SiLU \\
& Dense
& 128
& Linear \\
& $1\times$ Transformer
& 128
& SiLU \\
& Prediction Layer
& $|\mathcal V|$
& Softmax \\
\bottomrule
\end{tabular}
\end{table}

\textbf{Training Parameters.} The encoder and user-specific decoders are based on the Qwen2 Transformer~\cite{yang2024qwen2}, while the CSI-aware modulator employs a BiLSTM backbone~\cite{graves2005framewise}. The field-test
network configuration is summarized in
Table~\ref{tab:network_configuration}. The Transformer uses an
embedding dimension of 128, eight query heads, one key--value head, and an FFN dimension of 512. In simulations, the AP-symbol head outputs two real values representing the in-phase and quadrature components of a continuous complex symbol, whereas the field-test configuration uses $M_x$ output logits for discrete $M_x$-PSK signaling.

Each token is represented by $N_c=16$ complex channel uses, and the maximum sequence length is $L=30$. Unless otherwise specified, the models are trained with a batch size of 256 for 50 epochs and fine-tuned for an additional 5 epochs for field deployment. We use the Adam optimizer with $(\beta_1,\beta_2)=(0.9,0.98)$, $\epsilon=10^{-8}$, and weight decay $5\times10^{-4}$. The learning rate is set to $10^{-3}$, and the Gumbel--Softmax temperature is linearly annealed from 1 to 0.1 during training.

\textbf{Benchmarks.}
For semantic-reconstruction comparisons, we consider two separated
source--channel coding baselines:
\begin{itemize}
    \item \textit{Baseline I:} Huffman source
    coding~\cite{huffman1952method}, LDPC-coded $64$-QAM
    ~\cite{ryan2004introduction}, and $\Sigma\Delta$
    modulation~\cite{keung2024transmitting} for MTS-assisted
    multiuser transmission;
    
    \item \textit{Baseline II:} 5-bit fixed-length source
    coding~\cite{xie2021deep}, LDPC-coded $64$-QAM
    ~\cite{ryan2004introduction}, and $\Sigma\Delta$
    modulation~\cite{keung2024transmitting} for MTS-assisted
    multiuser transmission.
\end{itemize}
For our $20\times20$ planar MTS, the one-dimensional spatial
$\Sigma\Delta$ recursion is applied independently to each row, with the quantization-error state reset at the beginning of each row. For the field-test baselines, the estimated direct-link component \(\widehat d_k x_t\) is subtracted from the received signal before symbol detection. For a fair comparison, all schemes are allocated the same total number of transmitted complex symbols for each multiuser message group.

We further compare the effective throughput achieved by the proposed Semantic Prism with that of the following two Time Division Multiple Access (TDMA) schemes:
\begin{itemize}
    \item \textit{Conventional TDMA:} The method is the same as Baseline I, except that users are served sequentially. 
    \item \textit{Semantic TDMA:} The method uses the same semantic transceiver architecture as Semantic Prism with $K=1$, and users are served sequentially.
\end{itemize}
The effective throughput is defined as
\begin{equation}
R_{\rm eff}
=
\frac{N_{\rm word}^{\rm correct}}
{T_{\rm total}},
\end{equation}
where $N_{\rm word}^{\rm correct}$ denotes the total number of correctly recovered words, and $T_{\rm total}$ denotes the measured total payload-transmission time required to deliver the complete test set. For a fair comparison, all schemes transmit exactly the same set of sentences with the same total number of source words. For each scheme, $T_{\rm total}$ is the accumulated payload-transmission time required to deliver the complete test set.

\textbf{Simulation settings.} In our simulations, the AP transmits continuous-valued complex symbols, and the modulator is provided with perfect reflected-link CSI, while the direct AP--user links are blocked. The MTS retains the same discrete phase alphabet as in the prototype. We consider a single-RF-chain AP, a $20\times20$ planar MTS with $N=400$ meta-atoms, and $K\in\{2,\ldots,7\}$ users. The AP and MTS are located at $(10,-10,0)$ and $(0,0,0)$, respectively, while the users are independently and uniformly distributed over $x,y\in[5,15]$ with $z=-5$. The AP--MTS and MTS--user links follow a geometry-based Rician fading model, where the LoS components are determined by the corresponding array responses and the NLoS components are independently drawn from circularly symmetric complex Gaussian distributions. Large-scale attenuation follows the path-loss model $30+22\log_{10}(d)$~dB. Unless otherwise specified, the transmit power and noise power are set to $30$~dBm and $-80$~dBm, respectively.

\subsection{Modulator Architecture Ablation}
\label{subsec:modulator_ablation}

\begin{figure}[t]
    \centering
    \includegraphics[width=\linewidth]
    {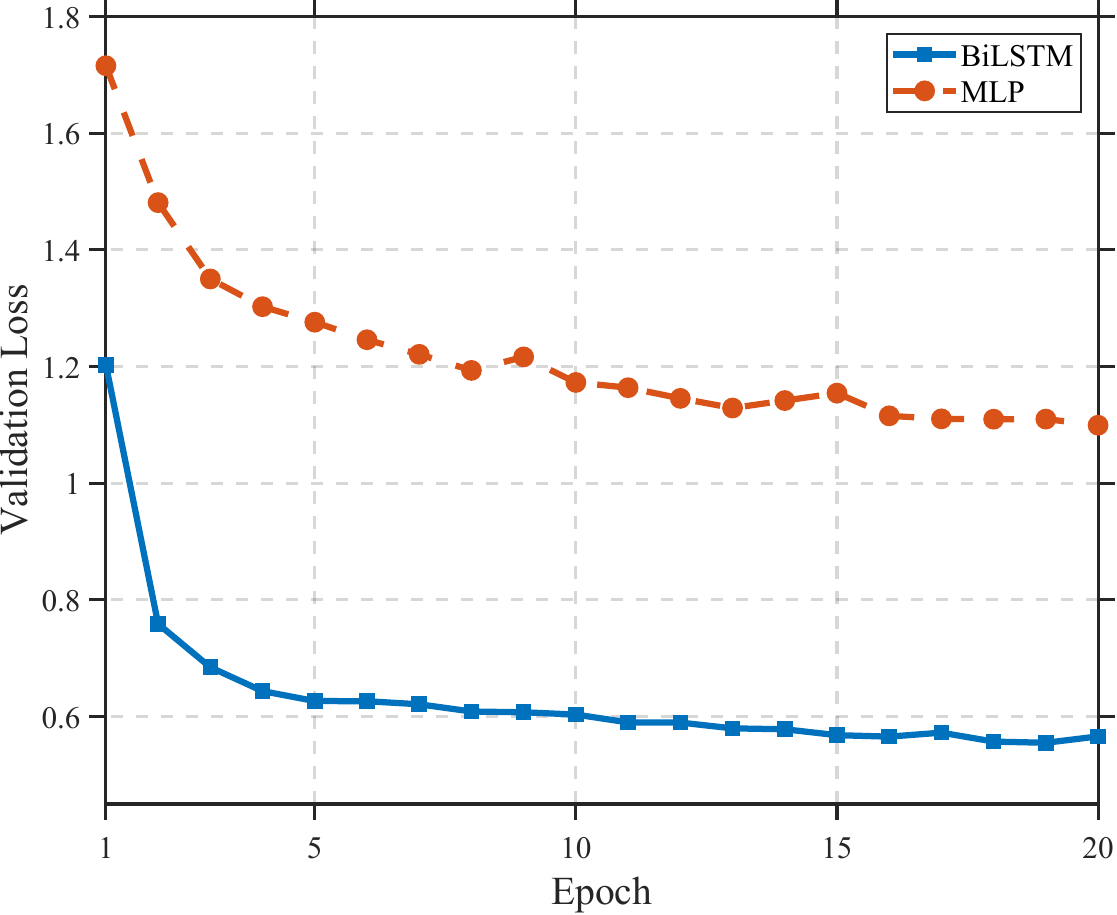}
    \caption{Validation-loss convergence of the BiLSTM-backbone and MLP-backbone variants of the CSI-aware modulator for $K=6$ under the idealized simulation setting.}
    \label{fig:bilstm_vs_mlp}
\end{figure}

We compare two implementations of the CSI-aware modulator, using either a BiLSTM or an MLP backbone. The two models are configured with comparable numbers of trainable parameters while sharing the same input features, output heads, and training settings. As shown in Fig.~\ref{fig:bilstm_vs_mlp}, the BiLSTM-based modulator achieves a substantially lower validation loss. This improvement can be attributed to its ability to capture correlations among semantic features across channel uses, thereby coordinating the AP symbols and MTS configurations over the transmission block and producing more coherent received feature sequences for semantic reconstruction. We therefore adopt the BiLSTM-based modulator in all subsequent experiments.

\subsection{BER Versus Semantic Fidelity}
\label{subsec:ber_to_semantics}
\begin{table}[t]
\centering
\caption{BER comparison under the idealized simulation setting with Rician factor $\kappa=10$, transmit power $P=30$~dBm, and
$N=400$ meta-atoms.}
\label{tab:huffman_ber_comparison}
\small
\setlength{\tabcolsep}{5pt}
\renewcommand{\arraystretch}{1.08}

\begin{tabular}{c c c}
\toprule
$K$
& Huffman--LDPC--$\Sigma\Delta$
& Semantic Prism \\
\midrule
2 & $3.7048\times10^{-4}$ & $7.7446\times10^{-2}$ \\
3 & $4.1514\times10^{-3}$ & $9.0218\times10^{-2}$ \\
4 & $1.5607\times10^{-2}$ & $9.6694\times10^{-2}$ \\
5 & $4.0530\times10^{-2}$ & $1.2529\times10^{-1}$ \\
6 & $8.0433\times10^{-2}$ & $1.4558\times10^{-1}$ \\
7 & $1.2157\times10^{-1}$ & $2.5281\times10^{-1}$ \\
\bottomrule
\end{tabular}
\end{table}

We examine whether bit error rate (BER) adequately reflects
text-reconstruction quality by comparing bit-level errors with semantic fidelity. As shown in Table~\ref{tab:huffman_ber_comparison}, the BER of the Huffman--LDPC--$\Sigma\Delta$ baseline increases from
$3.7048\times10^{-4}$ at $K=2$ to $1.2157\times10^{-1}$ at $K=7$,
reflecting the increasing difficulty of multiuser bitstream recovery. Semantic Prism has no native BER because it does not transmit an explicit source bitstream; we therefore compute a diagnostic BER by Huffman-encoding the source and reconstructed sentences with the same codebook and comparing the resulting bit sequences. The diagnostic BER is computed over the length of the source bit sequence, with any missing bits in a shorter reconstructed sequence counted as errors. Since Huffman coding is variable-length, a few word errors can shift subsequent bits and yield a large diagnostic BER. For example, at $K=7$, Semantic Prism has a diagnostic BER of $2.5281\times10^{-1}$ while maintaining a semantic similarity above $0.88$, whereas the separated baselines drop to approximately $0.40$ in Fig.~\ref{fig:similarity_users}.
We therefore use sentence-level semantic similarity~\cite{li2025semantic} as the primary metric and BLEU-1~\cite{bleu} as a secondary metric.

\subsection{Performance, Robustness, and Generalization}
\label{subsec:performance_robustness_generalization}

\begin{figure*}[t]
\centering
\subfigure[Semantic similarity]{
    \includegraphics[width=0.455\linewidth]
    {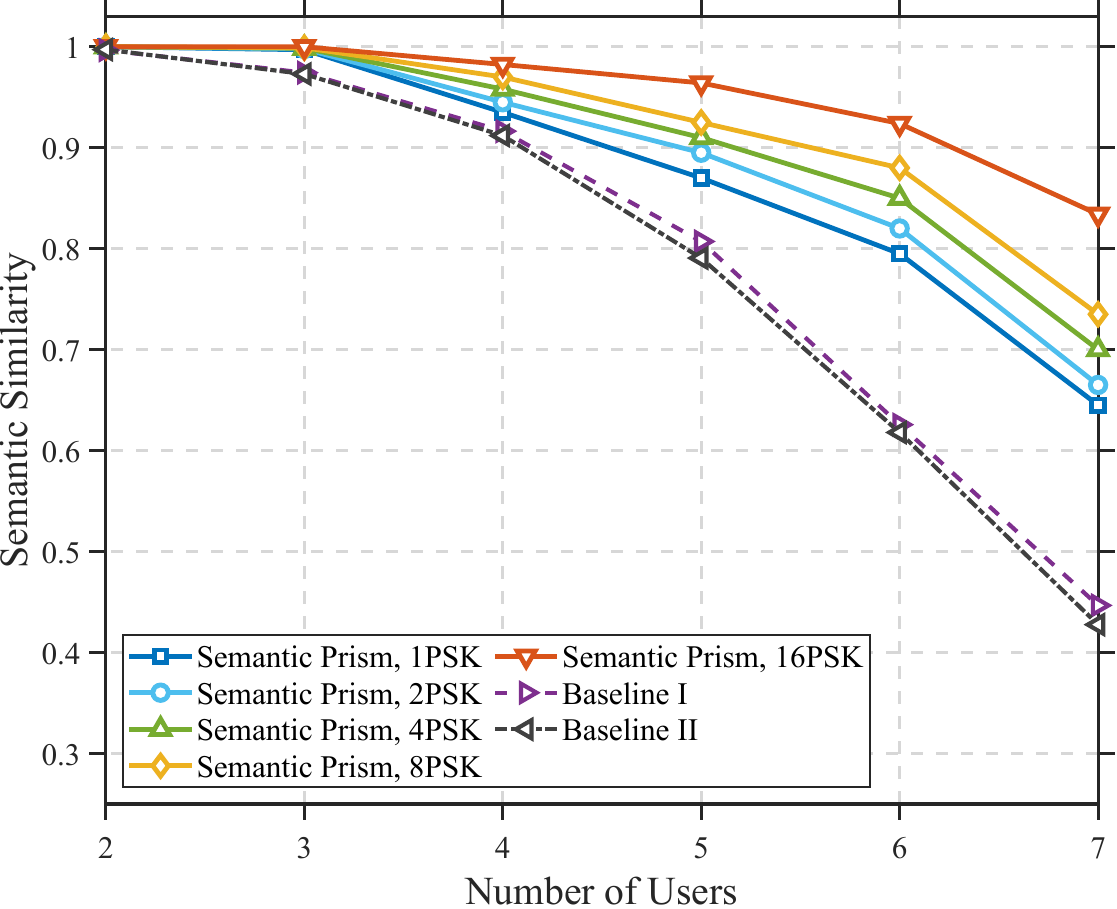}}
\hspace{1em}
\subfigure[BLEU-1]{
    \includegraphics[width=0.455\linewidth]
    {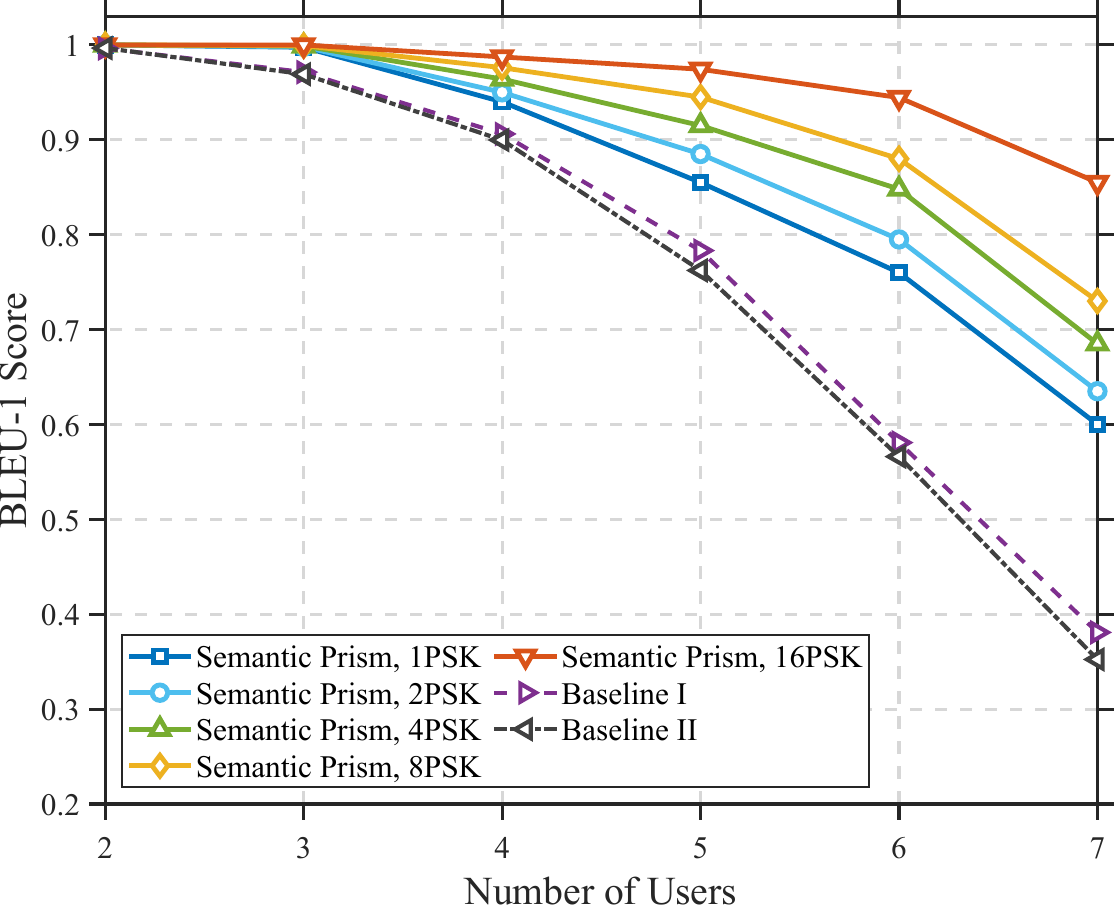}}
\caption{Reconstruction performance versus the number of users $K$ under different AP-side PSK orders in the field test. All proposed variants use the same 400-element MTS with 2-bit phase shifts.}
\label{fig:field_users_mpsk}
\end{figure*}

We first evaluate the multiuser reconstruction performance in the
field test. As shown in Fig.~\ref{fig:field_users_mpsk}, the
performance of all schemes decreases as $K$ increases because the
available spatial degrees of freedom must accommodate more simultaneous user-specific transmission targets. Nevertheless, the two separated baselines degrade rapidly when $K\geq4$, whereas Semantic Prism maintains considerably higher semantic similarity and BLEU-1 scores. Two practical factors contribute to this degradation. First, the baselines cancel the direct-link
component using its estimated coefficient $\widehat d_k$, and the
estimation error leaves a residual term $(d_k-\widehat d_k)x_t$ in the received signal, which degrades symbol detection. Second, the one-dimensional spatial $\Sigma\Delta$ recursion is applied independently to each row of the $20\times20$ MTS, with the quantization-error state reset at the beginning of each
row. Consequently, multiple boundary quantization-error terms remain across the rows, increasing the aggregate quantization distortion of the planar MTS. 

In contrast, Semantic Prism is more tolerant to these practical
impairments through both its transmission objective and physical-layer design. Rather than requiring exact recovery of prescribed modulation symbols, it reconstructs the intended messages from the received semantic-feature sequences, allowing moderate physical-layer distortion to be absorbed by the semantic decoder. Moreover, the CSI-aware modulator jointly designs the AP symbol and MTS configuration, providing additional flexibility beyond phase-only MTS modulation. For field deployment, the estimated direct-link CSI is explicitly incorporated into the modulator input, while channel perturbations are introduced during fine-tuning to improve robustness to CSI mismatch. These mechanisms make Semantic Prism robust to direct-link CSI mismatch while avoiding the row-wise $\Sigma\Delta$ quantization distortion that affects the separated baselines. Fig.~\ref{fig:field_users_mpsk} further shows that increasing the AP-side PSK order generally improves reconstruction performance, as a larger constellation provides more candidate AP symbols for jointly serving the users.




\begin{table*}[t]
\centering
\caption{Representative reconstruction for one user in a
$K=4$ multiuser transmission with 16-PSK AP signaling in the field test. Blue bold text marks meaning-preserving substitutions, while red bold text marks erroneous words or segments.}
\label{tab:semantic_synonym_balanced_20x20}
\small
\setlength{\tabcolsep}{6pt}
\renewcommand{\arraystretch}{1.18}
\begin{tabular}{p{0.15\linewidth} p{0.80\linewidth}}
\toprule
Version & Recovered text \\
\midrule

Original text &
One particularly praiseworthy aspect of this report is the attention it
pays to the promotion of equal opportunities for men and women in the
employment and labour markets. I believe the conclusions I have just
formulated very broadly answer the question raised by your rapporteur on
the interpretation of article. We afford you a measure of confidence and
support although we predict some very difficult negotiations ahead. \\
\midrule

Semantic Prism &
One particularly \textbf{\textcolor{blue}{inspiring}} aspect of this
report is the attention it pays to the promotion of equal opportunities
for men and women in the employment and labour markets. I believe the
conclusions I have just formulated very broadly answer the question
raised by your rapporteur on the interpretation of article. We afford
you a measure of confidence and support although we
\textbf{\textcolor{blue}{foresee}} some very difficult negotiations
ahead. \\
\midrule

Baseline I &
One \textbf{\textcolor{red}{palos ylarly}} praiseworthy aspect of
\textbf{\textcolor{red}{thidtaaertn sn oae}} attention it pays to the
promotion of equal \textbf{\textcolor{red}{opportunitrle ndr}} men and
women in the employment \textbf{\textcolor{red}{anm labnur marke}}.
I believe the \textbf{\textcolor{red}{concdusions}} I have just
\textbf{\textcolor{red}{formulatro}} very broadly answer the
\textbf{\textcolor{red}{kssseom e}} raised
\textbf{\textcolor{red}{byhetano pputosanofe ayiicaepea reoseo}} of
\textbf{\textcolor{red}{aertetoe}}.
We \textbf{\textcolor{red}{afxrd yow gmeasure}} of
\textbf{\textcolor{red}{confidencgand suspyto y naough}} we predict
some very \textbf{\textcolor{red}{difs s bdt}} negotiations ahead. \\
\midrule

Baseline II &
One \textbf{\textcolor{red}{particulatlz}} praiseworthy aspect of this
report is \textbf{\textcolor{red}{xhe astention}} it pays to the
\textbf{\textcolor{red}{ptomotion !f!hqual opportumitaes}} for men and
\textbf{\textcolor{red}{wkman}} in the
\textbf{\textcolor{red}{emplpymepd apd lajour}} markets.
\textbf{\textcolor{red}{M aelieve}} the
\textbf{\textcolor{red}{condlusions}} I have
\textbf{\textcolor{red}{justyfobmqlated}} very
\textbf{\textcolor{red}{broadg, bns!er tbe ruestion}} raised by your
rapporteur on the \textbf{\textcolor{red}{interoretation}} of
\textbf{\textcolor{red}{azticle}}.
We \textbf{\textcolor{red}{avfkrdwyou}} a measure of
\textbf{\textcolor{red}{confidefca}} and support although we
\textbf{\textcolor{red}{prelid. sommcvery tifficult neggtiations}}
ahead. \\
\bottomrule
\end{tabular}
\end{table*}

Table~\ref{tab:semantic_synonym_balanced_20x20} provides a representative text reconstruction for $K=4$ in the field test with 16-PSK AP signaling. Semantic Prism preserves most of the sentence structure and content, with only meaning-preserving lexical substitutions. For example, ``praiseworthy'' is replaced by ``inspiring,'' both conveying a positive evaluation in context, while ``predict'' is replaced by ``foresee,'' both expressing anticipation of a future event. In contrast, the separated baselines contain substantially more erroneous words and corrupted segments.  This example is consistent with the aggregate reconstruction results in Fig.~\ref{fig:field_users_mpsk}.

\begin{figure}[t]
    \centering
    \includegraphics[width=\linewidth]
    {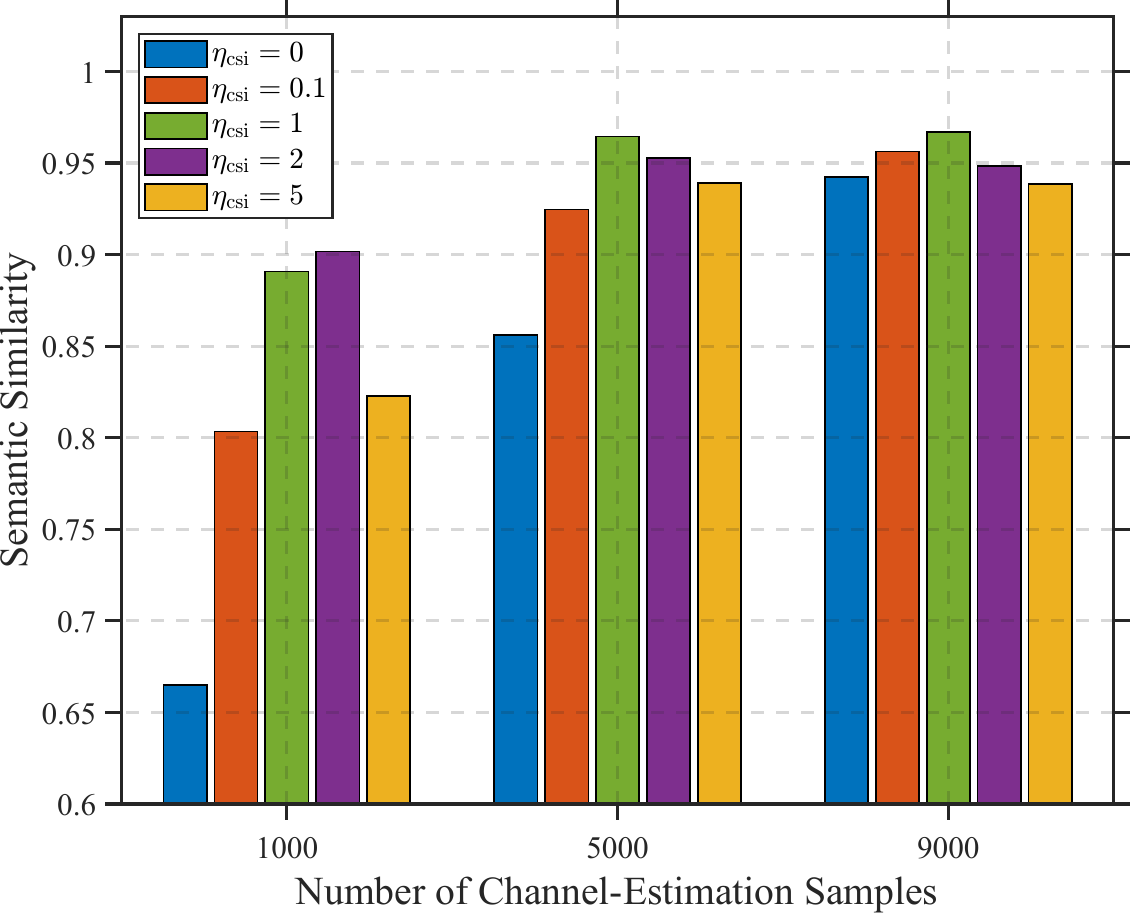}
    \caption{Semantic similarity versus the number of
    channel-estimation samples $T_{\rm ce}$ for $K=5$ in the field test. The case $\eta_{\rm csi}=0$ corresponds to fine-tuning without CSI perturbation, while $\eta_{\rm csi}>0$ corresponds to CSI-augmented fine-tuning.}
    \label{fig:field_estimation_samples}
\end{figure}

\begin{table}[t]
\centering
\caption{Channel-estimation overhead and semantic-similarity retention
for different values of $T_{\rm ce}$, normalized to those at
$T_{\rm ce}=9000$.}
\label{tab:estimation_sample_saving}
\footnotesize
\setlength{\tabcolsep}{5pt}
\renewcommand{\arraystretch}{1.1}
\begin{tabular}{c c c c}
\toprule
$T_{\rm ce}$
& CE overhead
& Reduction
& Similarity \\
\midrule
9000 & $100\%$ & --      & $100\%$ \\
5000 & $55.6\%$ & $44.4\%$ & $99.8\%$ \\
1000 & $11.1\%$ & $88.9\%$ & $93.1\%$ \\
\bottomrule
\end{tabular}
\end{table}

The robustness to channel-estimation errors is evaluated in
Fig.~\ref{fig:field_estimation_samples}. To account for inaccuracies in practical channel measurements, CSI perturbations are introduced during fine-tuning. Without such perturbations, the reconstruction performance becomes more sensitive as $T_{\rm ce}$ decreases, particularly at $T_{\rm ce}=1000$. In contrast, CSI-augmented fine-tuning improves the robustness to inaccurate channel estimates. As shown in Table~\ref{tab:estimation_sample_saving}, using $T_{\rm ce}=5000$ requires only $55.6\%$ of the channel-estimation overhead of $T_{\rm ce}=9000$ while retaining $99.8\%$ of its semantic
similarity. Further reducing $T_{\rm ce}$ to $1000$ lowers the overhead to $11.1\%$, with a similarity retention of $93.1\%$. We therefore use $T_{\rm ce}=5000$ as the default field-test setting.

\begin{figure}[t]
    \centering
    \includegraphics[width=\linewidth]
    {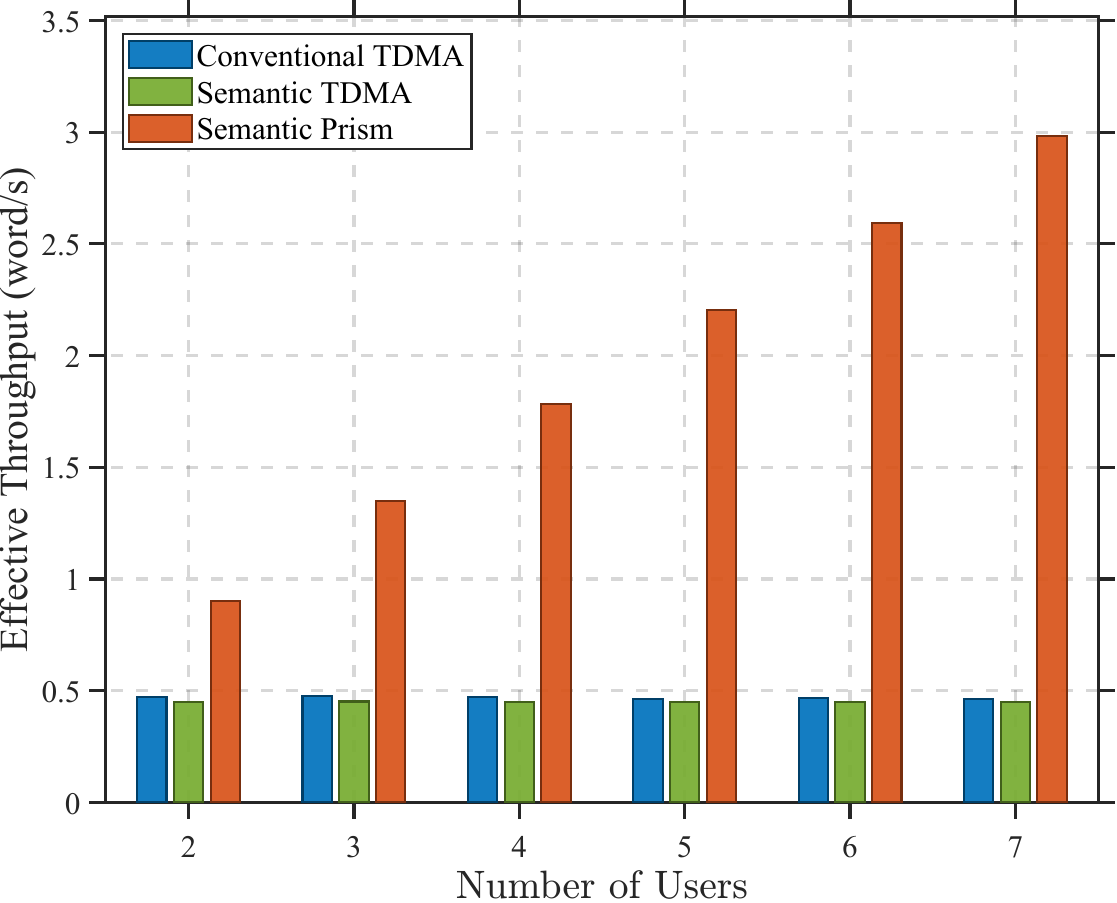}
    \caption{Effective throughput versus the
    number of users $K$ in the field test.}
    \label{fig:effective_throughput}
\end{figure}

The multiuser multiplexing benefit is further quantified by the
effective throughput in Fig.~\ref{fig:effective_throughput}. The TDMA
schemes serve users sequentially, whereas Semantic Prism exploits the
spatial degrees of freedom of the programmable MTS to concurrently
deliver independent messages with a single RF chain. By jointly
designing the AP symbol and MTS configuration, different
information-bearing effective signals are synthesized for different
users within the same transmission block. Consequently, more correctly recovered words can be delivered within the same transmission time as $K$ increases. The comparison with Semantic TDMA highlights this MTS-enabled multiplexing gain, since Semantic TDMA serves users sequentially, whereas Semantic Prism supports concurrent independent-message transmission.

\begin{figure}[t]
    \centering
    \includegraphics[width=\linewidth]
    {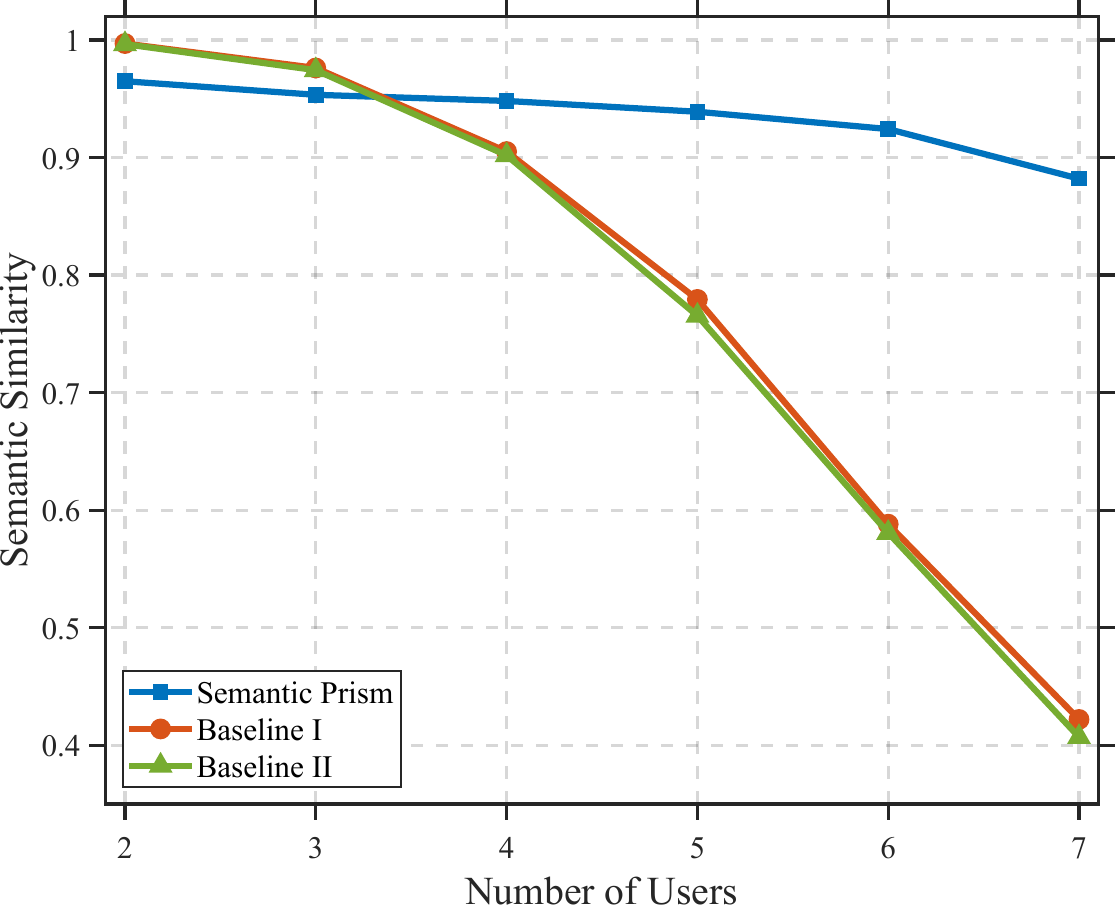}
    \caption{Semantic similarity versus the number of users $K$ under the simulation setting with Rician factor $\kappa=10$, $P=30$~dBm, and $N=400$.}
    \label{fig:similarity_users}
\end{figure}

\begin{figure}[t]
    \centering
    \includegraphics[width=\linewidth]
    {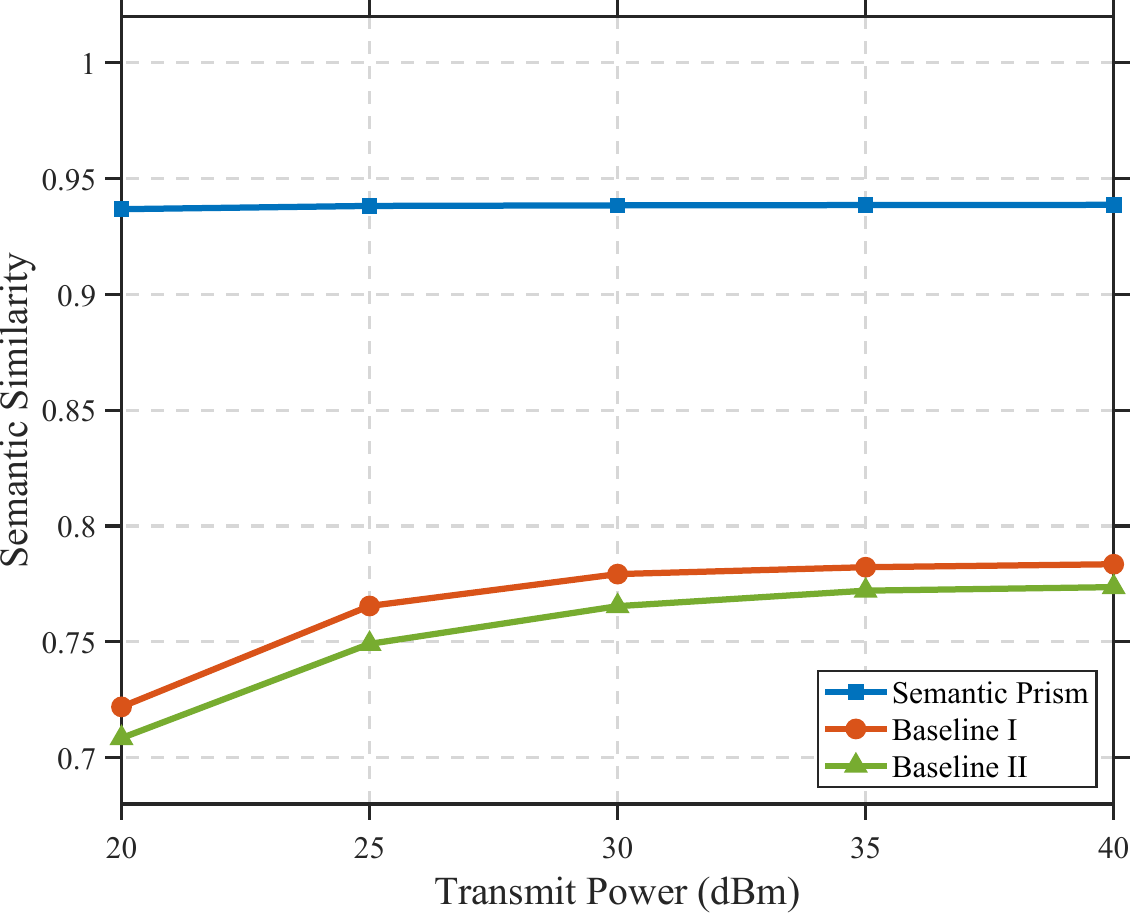}
    \caption{Semantic similarity versus transmit power $P$ under the simulation setting with Rician factor $\kappa=10$, $K=5$, and $N=400$.}
    \label{fig:similarity_power}
\end{figure}

\begin{figure}[t]
    \centering
    \includegraphics[width=\linewidth]
    {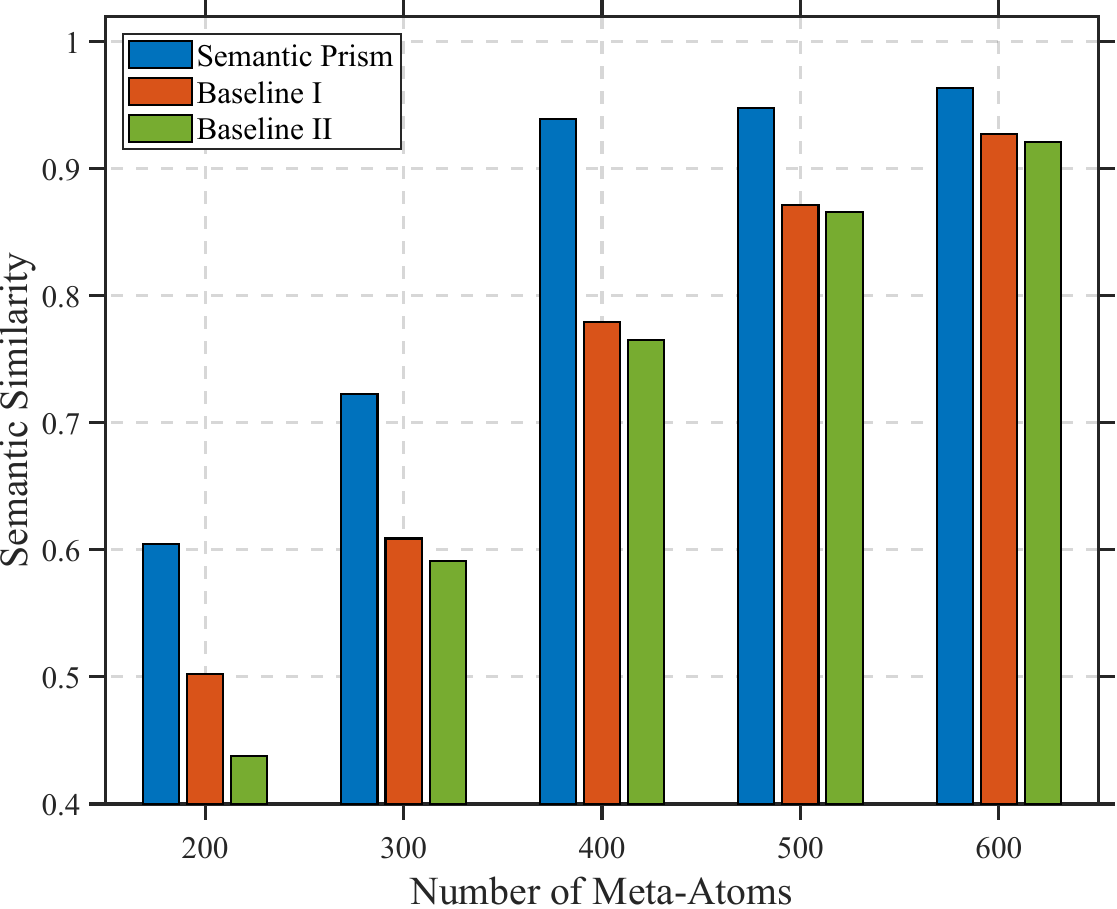}
    \caption{Semantic similarity versus the number of meta-atoms $N$
    under the simulation setting with Rician factor $\kappa=10$, $K=5$, and $P=30$~dBm.}
    \label{fig:similarity_elements}
\end{figure}

Additionally, we also evaluate the proposed method through simulations with continuous-valued AP symbols, perfect reflected-link CSI, and blocked direct AP--user links. Fig.~\ref{fig:similarity_users} shows that Semantic Prism maintains a clear advantage as the number of users increases. Fig.~\ref{fig:similarity_power} further shows that its
semantic similarity remains approximately $0.94$ over the considered
transmit-power range, whereas the two separated baselines remain below $0.80$ even at $40$~dBm. Increasing the MTS size improves all schemes, as shown in Fig.~\ref{fig:similarity_elements}. The advantage of Semantic Prism is more pronounced for smaller arrays; for example, at $N=200$, its semantic similarity is approximately $0.60$, compared with about $0.50$ and $0.44$ for Baselines I and II, respectively. The gap gradually narrows as more programmable meta-atoms become available.

\begin{figure}[t]
    \centering
    \includegraphics[width=\linewidth]
    {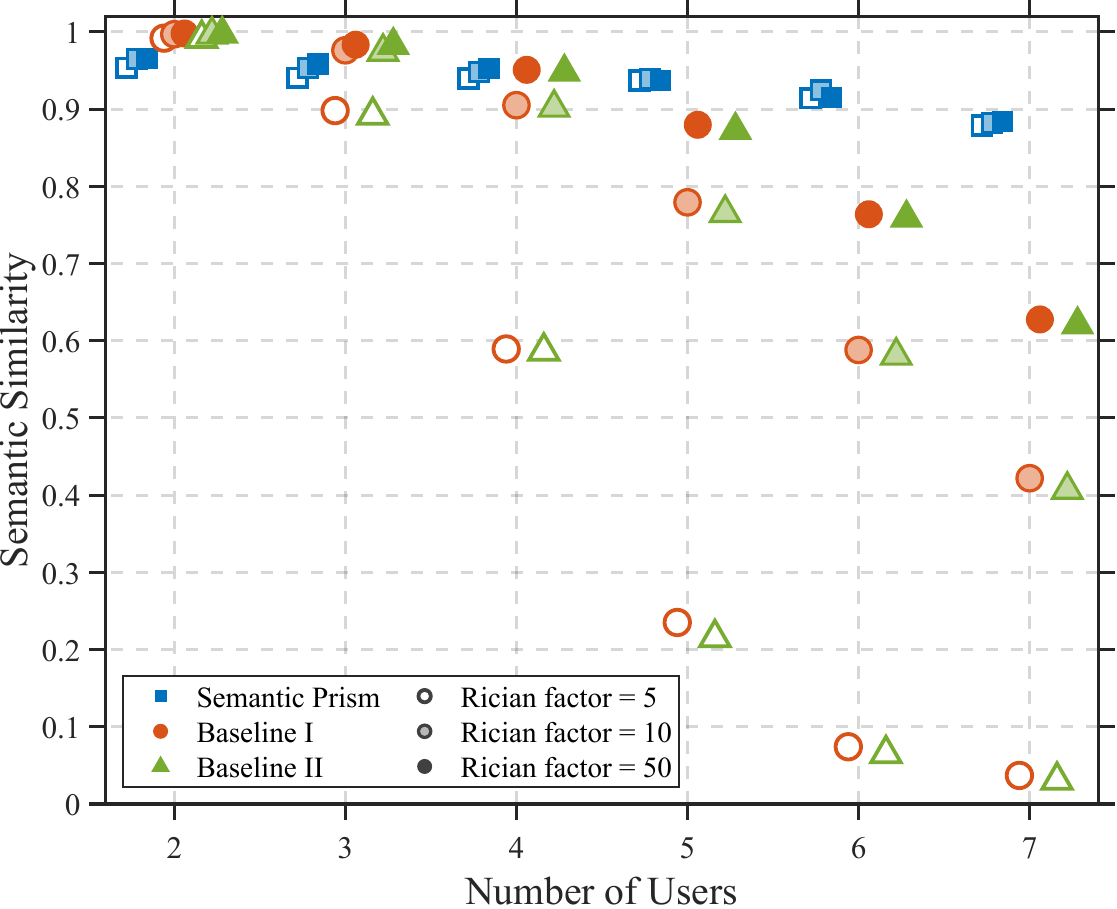}
    \caption{Semantic similarity versus the number of users $K$ under the simulation setting with different Rician factors.}
    \label{fig:generalization}
\end{figure}

Moreover, Fig.~\ref{fig:generalization} evaluates the generalization performance over different Rician factors. For each value of $K$,
the three markers correspond to Rician factors
$\kappa\in\{5,10,50\}$ and are slightly offset along the horizontal
axis for visualization. The Semantic Prism model is trained only under $\kappa=10$. As shown in the figure, Semantic Prism maintains
consistently high semantic similarity across the considered channel
conditions, indicating good generalization across different channel conditions. The separated baselines become more sensitive to increasing $K$ at smaller Rician factors, whereas their performance improves with $\kappa$ because a more LoS-dominant channel better matches the steering-vector structure exploited by spatial $\Sigma\Delta$ noise shaping. As scattering becomes stronger, this match weakens, reducing the effectiveness of quantization-error suppression.

\section{Conclusion}
\label{sec:conclusion}

This paper presents \emph{Semantic Prism}, an MTS-based framework for simultaneously transmitting independent semantic messages from a single-RF-chain AP to multiple users, thereby achieving semantic spatial multiplexing. Rather than using the MTS solely for channel enhancement, Semantic Prism jointly controls the AP symbols and discrete MTS configurations to convey user-specific semantic representations. To realize this joint control, we develop a CSI-aware learning architecture that maps multiuser semantic representations and CSI to the corresponding AP symbols and MTS configurations. To facilitate practical deployment, we further develop a field-channel fine-tuning method to mitigate the impact of CSI uncertainty. We implement Semantic Prism on a 3.5-GHz MTS prototype comprising 400 elements. Field-test results demonstrate that Semantic Prism supports concurrent semantic transmission to up to seven users while maintaining high text reconstruction quality.

\bibliographystyle{IEEEtran}
\bibliography{IEEEabrv,refs}

@STRING{IEEE_J_SP         = "{IEEE} Trans. Signal Process."}

@STRING{IEEE_J_JSAC       = "{IEEE} J. Sel. Areas Commun."}

@STRING{IEEE_J_COM        = "{IEEE} Trans. Commun."}

@STRING{IEEE_J_WCOM       = "{IEEE} Trans. Wireless Commun."}

@STRING{IEEE_J_WCOML      = "{IEEE} Wireless Commun. Lett."}

@STRING{IEEE_J_IT         = "{IEEE} Trans. Inf. Theory"}

@STRING{IEEE_J_IOT        = "{IEEE} Internet Things J."}

@STRING{IEEE_J_CCN        = "{IEEE} Trans. Cogn. Commun. Netw."}

@STRING{IEEE_M_COM        = "{IEEE} Commun. Mag."}

@inproceedings{bereyhi2020singleRF,
  author    = {Bereyhi, Ali and Jamali, Vahid and M{\"u}ller, Ralf R. and Tulino, Antonia M. and Fischer, Georg and Schober, Robert},
  title     = {A Single-{RF} Architecture for Multiuser Massive {MIMO} via Reflecting Surfaces},
  booktitle = {Proc. IEEE Int. Conf. Acoust., Speech Signal Process. (ICASSP)},
  pages     = {8688--8692},
  month     = may,
  year      = {2020},
  doi       = {10.1109/ICASSP40776.2020.9052989}
}

@article{li2021singleRF,
  author  = {Li, Qiang and Wen, Miaowen and Di Renzo, Marco},
  title   = {Single-{RF} {MIMO}: From Spatial Modulation to Metasurface-Based Modulation},
  journal = {IEEE Wireless Commun.},
  volume  = {28},
  number  = {4},
  pages   = {88--95},
  month   = aug,
  year    = {2021},
  doi     = {10.1109/MWC.021.2000376}
}

@article{yang2024qwen2,
  author  = {Yang, An and Yang, Baosong and Hui, Binyuan and Zheng, Bo
             and Yu, Bowen and Zhou, Chang and Li, Chengpeng and Li, Chengyuan
             and Liu, Dayiheng and Huang, Fei and others},
  title   = {{Qwen2} Technical Report},
  journal = {arXiv preprint arXiv:2407.10671},
  year    = {2024}
}

@article{graves2005framewise,
  author  = {Graves, Alex and Schmidhuber, J{\"u}rgen},
  title   = {Framewise Phoneme Classification with Bidirectional {LSTM} and Other Neural Network Architectures},
  journal = {Neural Netw.},
  volume  = {18},
  number  = {5--6},
  pages   = {602--610},
  year    = {2005},
  doi     = {10.1016/j.neunet.2005.06.042}
}

@inproceedings{keung2024transmitting,
  author    = {Keung, Wai-Yiu and Cheng, Hei Victor and Ma, Wing-Kin},
  title     = {Transmitting Data Through Reconfigurable Intelligent Surface: A Spatial Sigma--Delta Modulation Approach},
  booktitle = {Proc. IEEE Int. Conf. Acoust., Speech Signal Process. (ICASSP)},
  pages     = {9251--9255},
  month     = apr,
  year      = {2024},
  doi       = {10.1109/ICASSP48485.2024.10448246}
}

@article{xie2021deep,
  author  = {Xie, Huiqiang and Qin, Zhijin and Li, Geoffrey Ye and Juang, Biing-Hwang},
  title   = {Deep Learning Enabled Semantic Communication Systems},
  journal = IEEE_J_SP,
  volume  = {69},
  pages   = {2663--2675},
  year    = {2021},
  doi     = {10.1109/TSP.2021.3071210}
}

@article{li2025semantic,
  author  = {Li, Mingxiao and Shen, Kaiming and Cui, Shuguang},
  title   = {A Semantic Approach to Successive Interference Cancellation for Multiple Access Networks},
  journal = IEEE_J_IOT,
  volume  = {12},
  number  = {11},
  pages   = {16424--16437},
  month   = jun,
  year    = {2025},
  doi     = {10.1109/JIOT.2025.3532484}
}

@ARTICLE{huang2019RIS,
  author={Huang, Chongwen and Zappone, Alessio and Alexandropoulos, George C. and Debbah, Mérouane and Yuen, Chau},
  journal=IEEE_J_WCOM, 
  title={Reconfigurable Intelligent Surfaces for Energy Efficiency in Wireless Communication}, 
  year={2019},
  volume={18},
  number={8},
  pages={4157-4170}}

@ARTICLE{wu2020ris,
  author={Wu, Qingqing and Zhang, Rui},
  journal=IEEE_M_COM, 
  title={Towards Smart and Reconfigurable Environment: Intelligent Reflecting Surface Aided Wireless Network}, 
  year={2020},
  volume={58},
  number={1},
  pages={106-112}}

@article{roy2021ris,
  author  = {Karasik, Roy and Simeone, Osvaldo and Di Renzo, Marco and Shamai (Shitz), Shlomo},
  title   = {Adaptive Coding and Channel Shaping Through Reconfigurable Intelligent Surfaces: An Information-Theoretic Analysis},
  journal = IEEE_J_COM,
  volume  = {69},
  number  = {11},
  pages   = {7320--7334},
  month   = nov,
  year    = {2021},
  doi     = {10.1109/TCOMM.2021.3100621}
}

@article{yan2020ris,
  author  = {Yan, Wenjing and Yuan, Xiaojun and He, Zhen-Qing and Kuai, Xiaoyan},
  title   = {Passive Beamforming and Information Transfer Design for Reconfigurable Intelligent Surfaces Aided Multiuser {MIMO} Systems},
  journal = IEEE_J_JSAC,
  volume  = {38},
  number  = {8},
  pages   = {1793--1808},
  month   = aug,
  year    = {2020},
  doi     = {10.1109/JSAC.2020.3000811}
}

@inproceedings{cheng2022ris,
  author    = {Cheng, Hei Victor and Yu, Wei},
  title     = {Modulating Data Using Reconfigurable Intelligent Surface by Symbol-Level Precoding},
  booktitle = {Proc. Int. Symp. Wireless Commun. Syst. (ISWCS)},
  pages     = {1--6},
  year      = {2022},
  doi       = {10.1109/ISWCS56560.2022.9940262}
}

@ARTICLE{guo2020ris,
  author={Guo, Shuaishuai and Lv, Shuheng and Zhang, Haixia and Ye, Jia and Zhang, Peng},
  journal=IEEE_J_JSAC, 
  title={Reflecting Modulation}, 
  year={2020},
  volume={38},
  number={11},
  pages={2548-2561},
  doi={10.1109/JSAC.2020.3007060}}

@article{jiang2024ris,
  author  = {Jiang, Peiwen and Wen, Chao-Kai and Jin, Shi and Li, Geoffrey Ye},
  title   = {{RIS}-Enhanced Semantic Communications Adaptive to User Requirements},
  journal = IEEE_J_COM,
  volume  = {72},
  number  = {7},
  pages   = {4134--4148},
  month   = jul,
  year    = {2024},
  doi     = {10.1109/TCOMM.2024.3369697}
}

@article{cheng2023degree,
  author  = {Cheng, Hei Victor and Yu, Wei},
  title   = {Degree-of-Freedom of Modulating Information in the Phases of Reconfigurable Intelligent Surface},
  journal = IEEE_J_IT,
  volume  = {70},
  number  = {1},
  pages   = {170--188},
  month   = jan,
  year    = {2024},
  doi     = {10.1109/TIT.2023.3332425}
}

@article{xie2024star,
  author  = {Xie, Ping and Li, Fan and Zhang, Mingchuan and Quan, Wei and Zhu, Junlong and Cheng, Nan},
  title   = {{STAR-RIS} Assisted Information Transmission Based on Fairness in Semantic Communication Systems},
  journal = IEEE_J_WCOM,
  volume  = {23},
  number  = {11},
  pages   = {17007--17020},
  month   = nov,
  year    = {2024},
  doi     = {10.1109/TWC.2024.3449414}
}

@article{ma2024enhanced,
  author  = {Ma, Junyu and Li, Qiang and Liu, Ronghua and Pandharipande, Ashish and Ge, Xiaohu},
  title   = {Enhanced Semantic Information Transfer on {RIS}-Assisted Communication Systems},
  journal = IEEE_J_WCOML,
  volume  = {13},
  number  = {8},
  pages   = {2225--2229},
  month   = aug,
  year    = {2024},
  doi     = {10.1109/LWC.2024.3408420}
}

@inproceedings{vaswani2017attention,
  title={Attention is all you need},
  author={Vaswani, Ashish and Shazeer, Noam and Parmar, Niki and Uszkoreit, Jakob and Jones, Llion and Gomez, Aidan N and Kaiser, {\L}ukasz and Polosukhin, Illia},
  booktitle={Neural Info. Process. Syst. (NeurIPS)},
  volume={30},
  year={2017}
}

@article{huffman1952method,
  title={A method for the construction of minimum-redundancy codes},
  author={Huffman, David A},
  journal={Proc. Inst. Radio Eng.},
  volume={40},
  number= 9,
  month = {Sep,},
  pages={1098--1101},
  year={1952},
  publisher={IEEE}
}

@incollection{ryan2004introduction,
  author    = {Ryan, William E.},
  title     = {An Introduction to {LDPC} Codes},
  booktitle = {Coding and Signal Processing for Magnetic Recording Systems},
  editor    = {Vasic, Bane and Kurtas, Erozan M.},
  publisher = {CRC Press},
  address   = {Boca Raton, FL, USA},
  chapter   = {36},
  year      = {2004},
  doi       = {10.1201/9780203490310.ch36}
}

@InProceedings{bleu,
  author    = {Papineni, Kishore and Roukos, Salim and Ward, Todd and Zhu, Wei-Jing},
  title     = {{BLEU}: {A} method for automatic evaluation of machine translation},
  booktitle = {Proc. Annual Meeting Assoc. Comput. Linguistics (ACL’02)},
  year      = {2002},
  pages     = {311–318},
  
}

@inproceedings{koehn-2005-europarl,
  title={Europarl: A parallel corpus for statistical machine translation},
  author={Koehn, Philipp},
  booktitle={Proc. MT Summit},
  pages={79--86},
  month = {Sep,},
  year={2005}
}

@inproceedings{reimers2019sentence,
  author    = {Reimers, Nils and Gurevych, Iryna},
  title     = {Sentence-{BERT}: Sentence Embeddings Using Siamese {BERT}-Networks},
  booktitle = {Proc. Conf. Empirical Methods Natural Lang. Process. 
               Int. Joint Conf. Natural Lang. Process. (EMNLP-IJCNLP)},
  pages     = {3982--3992},
  month     = nov,
  year      = {2019},
  doi       = {10.18653/v1/D19-1410}
}

@inproceedings{rahutomo2012semantic,
  title={Semantic cosine similarity},
  author={Rahutomo, Faisal and Kitasuka, Teruaki and Aritsugi, Masayoshi and others},
  booktitle={Proc. 7th Int. Student Conf Adv. Sci. Technol.},
  volume={4},
  number={1},
  year={2012},
  organization={University of Seoul South Korea}
}

@inproceedings{crossword,
  author    = {Li, Mingxiao and Jin, Rui and Xiang, Liyao and Shen,Kaiming and Cui, Shuguang},
  title     = {CROSSWORD: A SEMANTIC APPROACH TO TEXT COMPRESSION VIA MASKING},
  booktitle = {Proc. IEEE Int. Conf. Acoust. Speech Signal Process. (ICASSP)},
  month = {April},
  year      = {2024},
  pages     = {9171-9175},
  
}

@article{wu2019intelligent,
  author  = {Qingqing Wu and Rui Zhang},
  title   = {Intelligent Reflecting Surface Enhanced Wireless Network via Joint Active and Passive Beamforming},
  journal = IEEE_J_WCOM,
  volume  = {18},
  number  = {11},
  pages   = {5394--5409},
  month   = nov,
  year    = {2019},
}

@inproceedings{jang2017categorical,
  author    = {Jang, Eric and Gu, Shixiang and Poole, Ben},
  title     = {Categorical Reparameterization with Gumbel--Softmax},
  booktitle = {Proc. Int. Conf. Learn. Represent. (ICLR)},
  year      = {2017}
}

@article{cui2019secure,
  author  = {Miao Cui and Guangchi Zhang and Rui Zhang},
  title   = {Secure Wireless Communication via Intelligent Reflecting Surface},
  journal = {IEEE Wireless Commun. Lett.},
  volume  = {8},
  number  = {5},
  pages   = {1410--1414},
  month   = oct,
  year    = {2019},
}

@article{weng2021semantic,
  author  = {Zhenzi Weng and Zhijin Qin},
  title   = {Semantic Communication Systems for Speech Transmission},
  journal = IEEE_J_JSAC,
  volume  = {39},
  number  = {8},
  pages   = {2434--2444},
  month   = aug,
  year    = {2021}
}

@article{bourtsoulatze2019deep,
  author  = {Eirina Bourtsoulatze and David Burth Kurka and Deniz G{\"u}nd{\"u}z},
  title   = {Deep Joint Source-Channel Coding for Wireless Image Transmission},
  journal = IEEE_J_CCN,
  volume  = {5},
  number  = {3},
  pages   = {567--579},
  month   = {Sept.},
  year    = {2019},
}

@article{xie2021lite,
  author  = {Huiqiang Xie and Zhijin Qin},
  title   = {A Lite Distributed Semantic Communication System for Internet of Things},
  journal = IEEE_J_JSAC,
  volume  = {39},
  number  = {1},
  pages   = {142--153},
  month   = jan,
  year    = {2021}
}

@article{zhang2024unified,
  author  = {Guangyi Zhang and Qiyu Hu and Zhijin Qin and Yunlong Cai and Guanding Yu and Xiaoming Tao},
  title   = {A Unified Multi-Task Semantic Communication System for Multimodal Data},
  journal = IEEE_J_COM,
  volume  = {72},
  number  = {7},
  pages   = {4101--4116},
  month   = jul,
  year    = {2024}
}

@article{xie2022task,
  author  = {Huiqiang Xie and Zhijin Qin and Xiaoming Tao and Khaled B. Letaief},
  title   = {Task-Oriented Multi-User Semantic Communications},
  journal = IEEE_J_JSAC,
  volume  = {40},
  number  = {9},
  pages   = {2584--2597},
  month   = {Sept.},
  year    = {2022}
}

@article{shi2023task,
  author  = {Yuanming Shi and Yong Zhou and Dingzhu Wen and Youlong Wu and Chunxiao Jiang and Khaled B. Letaief},
  title   = {Task-Oriented Communications for {6G}: Vision, Principles, and Technologies},
  journal = {IEEE Wireless Commun.},
  volume  = {30},
  number  = {3},
  pages   = {78--85},
  month   = jun,
  year    = {2023}
}

@inproceedings{chen2021pushing,
  title="Pushing the Physical Limits of {IoT} Devices with Programmable Metasurfaces",
  author={Chen, Lili and Hu, Wenjun and Jamieson, Kyle and Chen, Xiaojiang and Fang, Dingyi and Gummeson, Jeremy},
  booktitle="{USENIX} Symp. Netw. Sys. Design Implementation ({NSDI})",
  isbn = {978-1-939133-21-2},
  pages={425--438},
  year={2021},
  month = apr
}

@inproceedings{chen2023seamless,
author = {Chen, Lili and Yu, Bozhong and Ren, Ju and Gummeson, Jeremy and Zhang, Yaoxue},
title = {Towards Seamless Wireless Link Connection},
year = {2023},
booktitle = {Proc. Annu. Int. Conf. Mobile Syst., Appl., Services (MobiSys)},
pages = {137–149},
numpages = {13}
}

@techreport{3gpp38211,
  author      = {{3GPP}},
  title       = {{NR; Physical Channels and Modulation}},
  institution = {3rd Generation Partnership Project},
  number      = {TS 38.211},
  year        = {2021}
}

@article{xu2025chaotic,
  author  = {Xu, Jia Wen and Wei, Menglin and Zhang, Lei and Galdi, Vincenzo
             and Li, Lianlin and others},
  title   = {Chaotic Information Metasurface for Direct Physical-Layer
             Secure Communication},
  journal = {Nat. Commun.},
  volume  = {16},
  pages   = {5853},
  year    = {2025},
  doi     = {10.1038/s41467-025-60725-1}
}

@article{wang2025ris,
  author  = {Wang, Wen Hao and Wang, Jia Wei and Wan, Xiang and Wang, Xu Jie
             and Wan, Jia Chen and Li, Yue Heng and Zwick, Thomas
             and Wang, Cheng-Xiang and Cui, Tie Jun},
  title   = {A {RIS}-Aided {OFDM}-Based Integrated Sensing and Communication
             System for Communication, Detection, and Tracking},
  journal = {IEEE Trans. Veh. Technol.},
  volume  = {74},
  number  = {10},
  pages   = {15660--15673},
  year    = {2025},
  doi     = {10.1109/TVT.2025.3567106}
}

@article{yang2023energy,
  author  = {Yang, Zhaohui and Chen, Mingzhe and Zhang, Zhaoyang and Huang, Chongwen},
  title   = {Energy Efficient Semantic Communication Over Wireless Networks With Rate Splitting},
  journal = IEEE_J_JSAC,
  volume  = {41},
  number  = {5},
  pages   = {1484--1495},
  month   = may,
  year    = {2023},
  doi     = {10.1109/JSAC.2023.3240713}
}

@inproceedings{shi2023ris,
  author    = {Shi, Jiajia and Chan, Tse-Tin and Pan, Haoyuan and Lok, Tat-Ming},
  title     = {Reconfigurable Intelligent Surface Assisted Semantic Communication Systems},
  booktitle = {Proc. IEEE Int. Conf. Signal Process., Commun. Comput. (ICSPCC)},
  pages     = {1--6},
  year      = {2023},
  doi       = {10.1109/ICSPCC59353.2023.10400366}
}

@article{hu2024ris,
  author  = {Hu, Xuejie and Tian, Yue and Li, Qinying and Kho, Yau Hee
             and Wang, Xianling and Xiao, Baiyun and Yang, Zheng and Li, Wenda},
  title   = {A Novel {RIS}-Aided Optimization Strategy for Semantic Communication System},
  journal = IEEE_J_WCOML,
  volume  = {13},
  number  = {6},
  pages   = {1655--1659},
  month   = jun,
  year    = {2024},
  doi     = {10.1109/LWC.2024.3385492}
}
\end{document}